\documentclass[times,twocolumn,final,authoryear]{elsarticle}

\newcommand{\myfigref}[1]{Fig.~\ref{#1}}
\graphicspath{{./fig/}}
\usepackage{amsmath,amssymb} 
\usepackage[hidelinks]{hyperref}
\hypersetup{
    colorlinks=false,
}
\usepackage{multirow}

 \RequirePackage{geometry}
\usepackage{amssymb}
\usepackage{latexsym}

\usepackage{url}
\usepackage{xcolor}
\definecolor{newcolor}{rgb}{.8,.349,.1}

\begin{document}
\begin{frontmatter}

\title{Shape from Projections via Differentiable Forward Projector for Computed Tomography}

\author[1]{Jakeoung Koo\corref{cor1}} 
\cortext[cor1]{Corresponding author. E-mail: jakoo@dtu.dk\\
\textit{Accepted manuscript, Ultramicroscopy}\\
\textit{DOI: https://doi.org/10.1016/j.ultramic.2021.113239}\\
\textit{© 2021. This manuscript version is made available under the CC-BY-NC-ND 4.0 license http://creativecommons.org/licenses/by-nc-nd/4.0/}
}
\author[1]{Anders B. Dahl}
\author[1]{J. Andreas B{\ae}rentzen}
\author[2]{Qiongyang Chen}
\author[2]{Sara Bals}
\author[1]{Vedrana A. Dahl}

\address[1]{Technical University of Denmark, Anker Engelunds Vej 1, Kgs. Lyngby, 2800, Denmark}
\address[2]{University of Antwerp, Prinsstraat 13, Antwerpen, 2000, Belgium}


\begin{abstract}
In computed tomography, the reconstruction is typically obtained on a voxel grid. In this work, however, we propose a mesh-based reconstruction method. For tomographic problems, 3D meshes have mostly been studied to simulate data acquisition, but not for reconstruction, for which a 3D mesh means the inverse process of estimating shapes from projections. In this paper, we propose a differentiable forward model for 3D meshes that bridge the gap between the forward model for 3D surfaces and optimization. We view the forward projection as a rendering process, and make it differentiable by extending recent work in differentiable rendering. We use the proposed forward model to reconstruct 3D shapes directly from projections. Experimental results for single-object problems show that the proposed method outperforms traditional voxel-based methods on noisy simulated data. We also apply the proposed method on electron tomography images of nanoparticles to demonstrate the applicability of the method on real data. 
\end{abstract}


\end{frontmatter}


\section{Introduction}

In computed tomography (CT), we aim at solving the inverse problem of computing the 3D structure (shape and attenuation) of an object from a set of projection images~\citep{buzug2008computed} taken from different angles. Here, the geometry and the physics of the imaging system is known, which allows us to model the forward process, i.e.~if we have a suggestion for the 3D structure of the imaged object, we can compute the projection images.

We need a data structure to represent the structure of the object that should be reconstructed. The most common data structure is a volumetric image, with voxel intensities representing local attenuation. This approach may be used for reconstructing any type of object. Since each voxel in the volume is a parameter that must be computed, we have a very large number of unknowns. This makes it difficult to reconstruct volumes in situations where we have projections from a limited angular view (e.g., in electron tomography) or noisy data, and it can be difficult to accurately compute the attenuations in all voxels. Therefore, we propose to use a mesh to represent the shape of the object. The mesh separates the object into parts with a constant attenuation.

In tomography, forward projection of 3D meshes has mostly been used for simulating tomographic data acquisition, i.e.\ modeling the forward projection. This includes modeling X-ray transmission imaging based on Monte-Carlo methods~\citep{bonin2002montecarlo,freud2006fast} or ray tracing techniques~\citep{freud2006fast,marinovszki2018efficient}. Furthermore, \citet{vidal2009simulation,sujar2017gvirtualxray} took the advantage of the OpenGL library to simulate X-ray images in real time. However, none of the proposed methods are concerned with reconstruction, i.e. solving the inverse problem.

\begin{figure}[t]
    \centering
    \includegraphics[totalheight=100pt]{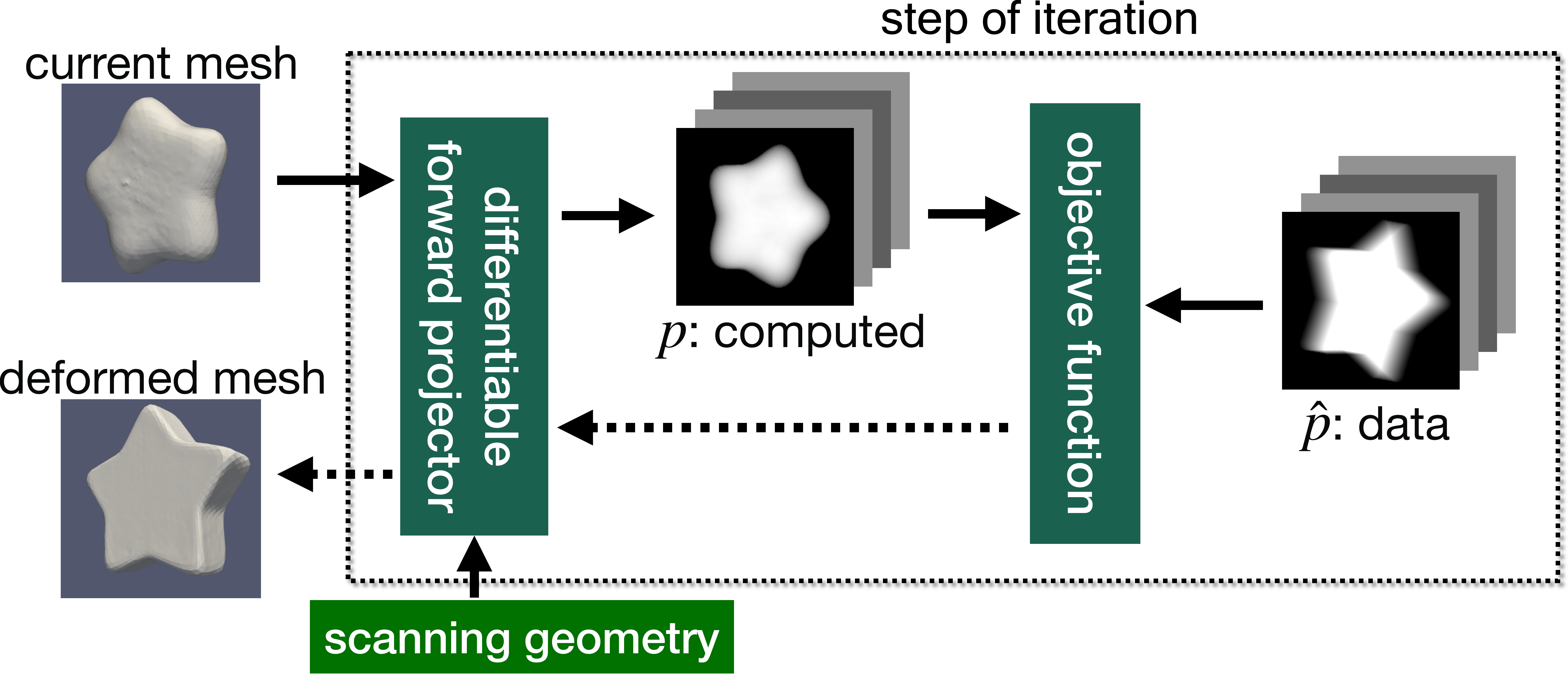}
	\caption{The proposed differentiable forward projector enables optimizing a 3D mesh from projections for tomographic reconstruction. Given a mesh and scanning geometry, the forward projector computes projections for the current mesh, which is updated by deforming vertices in the direction that minimizes our objective function.}
\label{fig:fig0}
\end{figure}

\begin{figure}[t]
    \centering
    \includegraphics[totalheight=64pt]{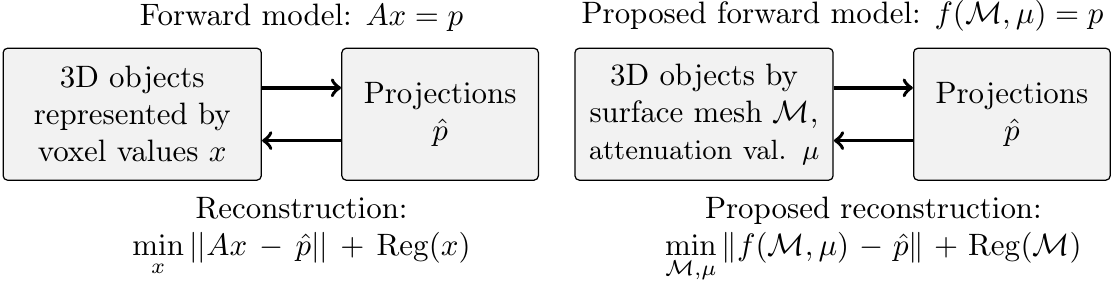}
	\caption{Algebraic reconstruction vs. our proposed approach. \emph{Left:} Existing algebraic reconstruction methods use a linear system of equations to model the forward projections typically together with some regularization to constrain the solution. \emph{Right:} We propose a differentiable forward model $f$ which computes projections for an object represented by a triangular mesh and attenuation coefficients $\mu$. This forward model is used to reconstruct the shape from projections data $\hat p$.}
\label{fig:fig1}
\end{figure}

For the mesh-based tomographic reconstruction that we propose, the reconstruction problem is two-fold. The mesh must be deformed to follow the boundaries of the depicted object, and in each part of the object, a single attenuation coefficient must be estimated. To forward project the mesh, we employ rendering techniques, which allow a very efficient projection of the 3D mesh to the detector plane. We extend the differentiable rasterizer recently proposed by~\citet{chen2019learning} to derive differentiable forward projection. This enables us to compute vertex displacements that deform the mesh based on the difference between the forward projection and projection data. In Fig.~\ref{fig:fig0}, we provide an overview of our shape estimation method. In Fig.~\ref{fig:fig1}, we illustrate how our work differs from existing image-based reconstruction methods~\citep{andersen1984simultaneous,buzug2008computed}.

Our model has been developed for problems like X-ray CT where the damping of the X-ray attenuation coefficient are modeled as linearly dependent on the path length through the sample~\citep{buzug2008computed}, which is also assumed in electron tomography up to a certain thickness~\citep{midgley20033d} where coefficient is related to electron scattering. 

In summary, our contribution is two-fold. We suggest a differentiable forward projector to generate projections from 3D meshes, and we propose a shape estimation method employing the differentiable forward projector. Our code is available online: \href{https://github.com/jakeoung/ShapeFromProjections}{https://github.com/jakeoung/ShapeFromProjections}.

\section{Related works}

\paragraph{Differentiable rendering based on rasterization} 
In our work we deform a mesh by changing vertex positions using gradient descent. Therefore, we need the gradient of our objective function with respect to vertex coordinates, which is possible through differentiating the forward projection. As observed in~\citep{vidal2009simulation}, the tomographic forward projection can be implemented by extending a rendering technique in computer graphics based on rasterization. Such rasterization-based rendering technique projects 3D models onto 2D image plane and involves a discrete step to choose the pixels covered by triangles of 3D models. This discrete step is not differentiable in conventional graphics pipelines. Making such rasterization-based rendering differentiable, called \textit{differentiable rasterization}, has been studied by several works to connect rendering and optimization. As our work extends such differentiable rasterization techniques, we first review those works and explain the difference to our work.

The general framework OpenDR, that was proposed by \cite{loper2014opendr}, approximated gradients of pixel values with respect to model parameters. \cite{kato2018neural} suggested a heuristic forward and backward pass where blurring is used to avoid zero gradients. This approach has an inconsistency between the forward and backward pass, and to circumvent this inconsistency, \cite{liu2019soft} proposed a method called SoftRas by relaxing discrete rasterization process into the aggregation of smooth probability functions. Unlike the conventional rasterization rendering, in SoftRas, each face in the mesh affects many pixels in the image plane, which is computationally costly and memory-demanding.

\cite{chen2019learning} suggested an interpolation-based differentiable rasterizer called DIB-R. DIB-R reformulates the barycentric interpolation in the rasterization process to analytically derive the gradients. Our forward projector extends this reformulation when computing the thickness of an object for computed tomography, but differs in some aspects. The rendering techniques such as DIB-R can improve the performance by ignoring invisible faces, but since we model penetrating radiation, our forward projector needs to consider all the faces. DIB-R used the idea of SoftRas~\citep{liu2019soft} for background pixels to propagate the gradients on those background, but our forward projector does not use it to reduce the computational cost.

\paragraph{Shape reconstruction from projections} 
Our proposed method is related to tomographic segmentation, where segments are directly computed from projections. This includes  ~\citep{elangovan2001sinograms,whitaker2002direct,alvino2004tomographic} that are based on the Mumford-Shah model~\citep{mumford1989optimal} where boundaries are represented using level-sets~\citep{osher2004level}. Recently, the parametric level-set method~\citep{aghasi2011parametric} has been used for tomographic segmentation in~\citep{kadu2018parametric,eliasof2020multimodal} where level-sets are represented as an aggregation of radial basis functions. Although the parametric level-set method has fewer unknown variables, its forward projection still depends on a regular grid.  \cite{gadelha2019shape} used a deep convolutional neural network for 2D tomographic reconstruction, where the forward projection is based on the transformation of a regular grid and resampling.
On the other hand, the work~\citep{dahl2018computing} based on snakes~\citep{kass1988snakes} avoids a dense grid -- it represents curves explicitly and proposes a direct forward projection of the curves. However, this method is limited to a single 2D curve, while the proposed method supports 3D objects. Another difference is that \citep{dahl2018computing} evolves curves in the normal directions of curve points, while our deformation can displace the vertices in all directions. 

In summary, existing shape estimation methods from projections are either based on regular grid or limited to single 2D curve. To our knowledge, our work is the first to propose the differentiable forward projector for 3D triangular mesh and use it for reconstructing shapes from projections.

\section{Differentiable forward projector}

In this section, we describe our main contribution of the differentiable forward projector. The goal is to forward project triangular meshes and make this process differentiable with respect to 3D vertex positions and attenuation coefficients. This differentiable forward projection will be used for optimizing the mesh shape described in Sec.~\ref{sec:shape}. First we describe the case of a single object and then extend to composite objects.

\subsection{Single object} \label{sec:single}

Consider an object represented by a watertight triangular mesh. A watertight triangle mesh forms a closed surface that has a well-defined interior and exterior: any path from a point in the interior to a point in the exterior must cross the triangle mesh.
We assume that the object is homogeneous, i.e. it has a certain attenuation coefficient $\mu$ associated with the volume inside the mesh. For now, we consider $\mu$ constant, we will later explain how it is computed in the next subsection.  The mesh consists of $K$ vertices, and we write $\mathbf{v}_k$ for the 3D coordinates of the vertex $k$. 

\begin{figure}[t]
    \centering
    \includegraphics[totalheight=130pt]{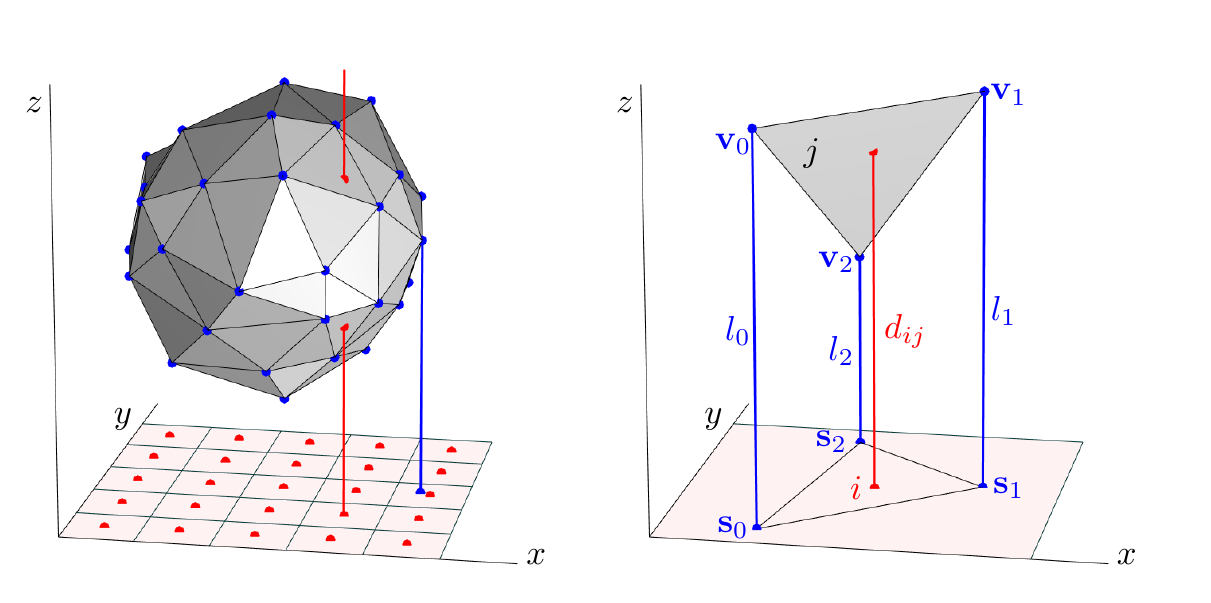}
	\caption{\emph{Left:} The vertices of the triangle mesh (blue dots) are projected onto the detector. Each detector pixel (red dots) is associated with the projection ray which intersects mesh triangles. \emph{Right:} One triangle $j$, here given by  vertices $k=0,1,2$, and one detector pixel $i$. Using barycentric coordinates, the distance $d_{ij}$ may be expressed in terms of $l_k$.}
\label{fig:forward}
\end{figure}

To simplify the explanation, we start with a single projection and later expand to multiple projections from different angles. Let $\mathbf P$ and $\mathbf R$ be the position of the detector and a matrix that rotates from detector coordinates to the global frame, respectively. If we denote the position of vertex $k$ in global coordinates by $\mathbf V_k$, the position in detector coordinates are $\mathbf v_k = \mathbf R^\intercal (\mathbf V_k - \mathbf P)$. Note that in detector coordinates, the detector itself corresponds to the plane $z=0$, its center is at the origin, and the positive $z$-axis points towards the object, see \myfigref{fig:forward}. 

Expressed in detector coordinates, the distance of the vertex $k$ from the detector is trivially $l_k = \mathbf{e}_3^\intercal \mathbf{v}_k$ while $\mathbf{s}_k = [\mathbf{e}_1\, \mathbf{e}_2]^\intercal \mathbf{v}_k$ are the coordinates of the projection of the vertex onto the detector. Here $\mathbf{e}_1$, $\mathbf{e}_2$, and $\mathbf{e}_3$ are unit vectors in $x$, $y$ and $z$ direction, for example $\mathbf{e}_1 = [1\,0\,0]^\intercal$. 

Projecting the object onto the detector pixel $i$ we consider the projection ray associated with $i$ (slightly sloppy, call it ray $i$), and its path length in the object. As explained in~\citep{vidal2009simulation}, this can be broken into contribution of all intersections of the ray $i$ with the mesh triangles
\begin{equation} \label{eq:pi}
p_i = \mu \sum_{\substack{j\\i \text{ intersects } j}} \operatorname{sign} (\mathbf{e}_3^\intercal \mathbf{n}_j) d_{ij}
\end{equation}
where $\mathbf{n}_j$ is the normal of the triangle $j$ (needed for determining the sign of the contribution), and $d_{ij}$ is the distance of the intersection point to the detector. Here, we consider the sign value $\operatorname{sign} (\mathbf{e}_3^\intercal \mathbf{n}_j)$ as a constant attribute of each triangle.

Considering now a single triangle $j$ we express $d_{ij}$ using barycentric coordinates
\begin{equation}
d_{ij} =  \sum_{\substack{k \\ k \text{ in } j}} w_{ij}^k l_k  \enspace ,
\end{equation}
where $k$ are the indices of the three vertices of the triangle $j$ and $w_{ij}^k$ are the corresponding three barycentric coordinates of pixel $i$ with respect to the projection of triangle $j$ onto the detector plane, see \myfigref{fig:forward}, \emph{right}. 

To make the forward projection differentiable, we derive
\begin{equation} \label{eq:dpdvk}
\frac{\partial p_i}{\partial \mathbf{v}_k} = \mu \sum_{\substack{j\\i \text{ intersects } j}} \operatorname{sign} (\mathbf{e}_3^\intercal \mathbf{n}_j) \frac{\partial d_{ij}}{\partial \mathbf{v}_k} \enspace ,
\end{equation}
and
\begin{equation} \label{eq:dddv}
\frac{\partial d_{ij}}{\partial \mathbf{v}_k} =  \sum_{\substack{k \\ k \text{ in } j}}  \left(w_{ij}^k \frac{\partial l_k}{\partial \mathbf{v}_k} + \frac{\partial w_{ij}^k}{\partial \mathbf{v}_k}  l_k\right) =  \sum_{\substack{k \\ k \text{ in } j}}  \left([0\,0\,w_{ij}^k] + \frac{\partial w_{ij}^k}{\partial \mathbf{v}_k}  l_k\right) .
\end{equation}
For the last step, computation of $\partial w_{ij}^k/\partial \mathbf{v}_{k}$, we employ the idea from~\citep{chen2019learning}, which reformulates the barycentric form to express the coefficients $w_{ij}^k$ in terms of 2D projected positions $\mathbf{s}_k$ and the position of detector pixel $i$.

\subsection{Composite objects} \label{sec:composite}
The method described above generalizes to composite objects if certain conditions are met. Specifically, we require that we know beforehand the topology of the parts of the composites, and how parts are embedded within one another. Thus, for each interface triangle we will have a suggestion that what class of material is on either side, but we do not know the specific attenuation coefficient of the classes, since we solve for those.

In order to extend Eq.~\eqref{eq:pi} to composites, we only have to observe that triangles may now be the interface between two materials and not just air and material. This can be handled simply by letting each triangle contribute twice
\begin{eqnarray} \label{eq:multiple}
p_i &=& \sum_{j} \mu_{j}  \operatorname{sign} (\mathbf{e}_3 ^\intercal \mathbf n_j ) d_{ij} - \sum_{j} \overline\mu_{j}  \operatorname{sign} (\mathbf{e}_3 ^\intercal \mathbf n_j ) d_{ij} \\
&=&\sum_{j} ( \mu_{j} - \overline\mu_{j} ) \operatorname{sign} (\mathbf{e}_3 ^\intercal \mathbf n_j ) d_{ij} \enspace ,
\end{eqnarray}
where $\mu_j$ is the attenuation of the interior material and $\overline\mu_j$ of the exterior material according to normal orientation. Of course, either attenuation will be zero if the material on the corresponding side of the triangle is air.

The derivative of a pixel value with respect to the contributing attenuation $\mu_m$ for material $m$ (by abuse of notation) is
\begin{equation} \label{eq:dpdm}
\frac{\partial p_{i}}{\partial \mu_{m}} = \sum_j \pm \operatorname{sign} (\mathbf{e}_3 ^\intercal \mathbf n_j ) d_{ij} \enspace,
\end{equation}
%
where the $\pm$ is positive if the interior material of face $j$ is labeled as $m$ and negative if $m$ is the exterior material. We also modify Eq.~\eqref{eq:dpdvk} by changing $\mu$ to $(\mu_j - \overline \mu_j)$.

We have derived the Jacobians in Eq.~\eqref{eq:dpdvk} and \eqref{eq:dpdm}, which will be used to optimize an objective function $E$. That is, we can propagate the gradients from $E$
\begin{align} \label{eq:dEdv}
\frac{\partial E}{\partial \mathbf v_{k}} = \sum_{i} \frac{\partial E}{\partial p_{i}} \frac{\partial p_i}{\partial \mathbf{v}_k}, 
\quad\quad\quad \frac{\partial E}{\partial \mu_{m}} = \sum_{i} \frac{\partial E}{\partial p_{i}} \frac{\partial p_i}{\partial \mu_{m}},
\end{align}
where the summation is over all the detector pixels. 

\section{Shape from projections} \label{sec:shape}
In this section, we use the proposed forward projector to reconstruct shapes from projections. We assume that a template mesh with the correct topology is given. We aim to deform the template mesh and estimate the attenuation coefficients by minimizing the residual between data $\hat p$ and our estimation $p$.

Optimizing only the data fitting term can lead to degenerate meshes. To obtain high-quality meshes, we impose three regularization terms. The first term is the Laplacian term~\citep{wang2018pixel2mesh}, which constraints the vertices to move similarly with their neighbors, defined by
\begin{equation}
E_\mathrm{lap}=\sum_{k} \| \mathbf V_{k} - \frac{1}{| \mathcal N(k) |}\sum_{n \in \mathcal N(k)} \mathbf V_{n}\|^2,
\end{equation}
where $\mathcal N(k)$ is the index set of neighboring vertices to $k$-th vertex.
The second term is the edge length term to penalize long edges
\begin{equation}
E_\mathrm{edge} = \sum_{(\mathbf V_{k}, \mathbf V_{n}) \in G} \| \mathbf V_{k} - \mathbf V_{n} \|^{2},
\end{equation}
where $G$ denotes the set of edges. Lastly, we impose the flattening term~\citep{kato2018neural,liu2019soft}
\begin{equation}
E_\mathrm{flat}=\sum_{e \in G}(1 - \cos \theta_{e})^{2},
\end{equation}
where $\theta_{e}$ is the angle between the normal vectors of two faces sharing the edge $e$. Flattening term is needed to remove near-zero volume spikes. These thin artifacts have negligible contribution to the forward projection, and will be ignored by the data fitting term. As shown in Fig.~\ref{fig:deform}, such artifacts can appear during the deformation, but disappear later.

With the data fidelity and regularization terms, the objective function to minimize is
\begin{equation} \label{eq:energy}
E(\{\mathbf V_{k}\}, \{\mu_{m}\}) = \| p - \hat p\|_{2}^{2} + \alpha E_\mathrm{lap} + \beta E_\mathrm{edge} + \gamma E_\mathrm{flat}
\end{equation}
where $\alpha, \beta, \gamma$ control the relative weights between the terms. Note that the size of projection data $\hat p$ is the number of detector pixels times the number of projection angles. We use automatic differentiation to minimize $E$. For large data, we can use stochastic gradient descent with mini batches in terms of projection angles. In this paper, however, we only consider full batch size of data.


\def\fw{42pt}
\def\fh{48pt}
\begin{figure}[t]
\begin{center}
\begin{tabular}{c@{ }c@{ }c@{ }c@{ }c}
\parbox[b][\fh][c]{\fw}{GT} & 
\includegraphics[totalheight=\fh]{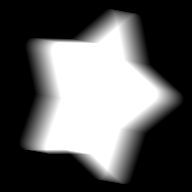} & \includegraphics[totalheight=\fh]{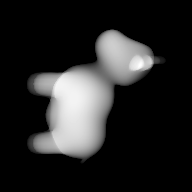} & \includegraphics[totalheight=\fh]{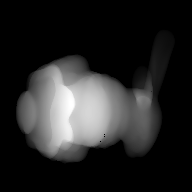} & \includegraphics[totalheight=\fh]{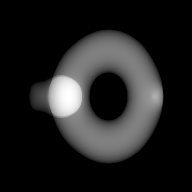} \\
\parbox[b][\fh][c]{\fw}{SIRT} & 
\includegraphics[totalheight=\fh]{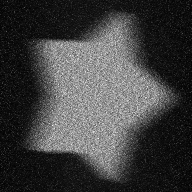} & \includegraphics[totalheight=\fh]{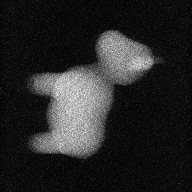} & \includegraphics[totalheight=\fh]{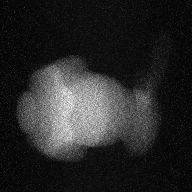} &
\includegraphics[totalheight=\fh]{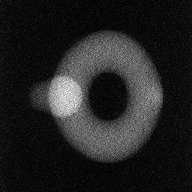} \\
\parbox[b][\fh][c]{\fw}{TV} & 
\includegraphics[totalheight=\fh]{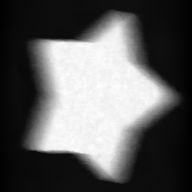} & \includegraphics[totalheight=\fh]{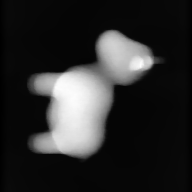} & \includegraphics[totalheight=\fh]{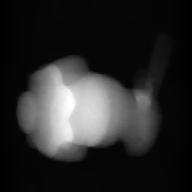} &
\includegraphics[totalheight=\fh]{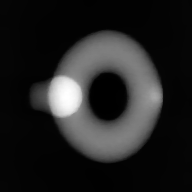}  \\
\parbox[b][\fh][c]{\fw}{Proposed} & 
\includegraphics[totalheight=\fh]{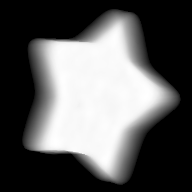} & \includegraphics[totalheight=\fh]{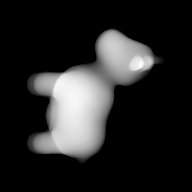} & \includegraphics[totalheight=\fh]{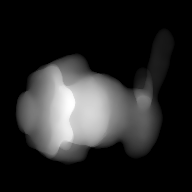} &
\includegraphics[totalheight=\fh]{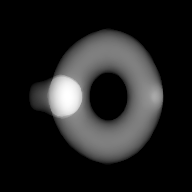}  \\
\end{tabular}
\caption{Qualitative results of estimated projections on noisy data with relative noise level 0.4. The first row shows the ground truth, \emph{i.e.}, noise-free data. 2nd-5th row show the forward projection from the solutions by SIRT, TV and the proposed method, respectively.}
\label{fig:projection}
\end{center}
\end{figure}

\def\fh{47pt}
\begin{figure}[!ht]
\centering
\begin{tabular}{c@{ }c@{ }c@{ }c@{ }c@{ }c@{ }c}
\includegraphics[width=\fh,totalheight=\fh]{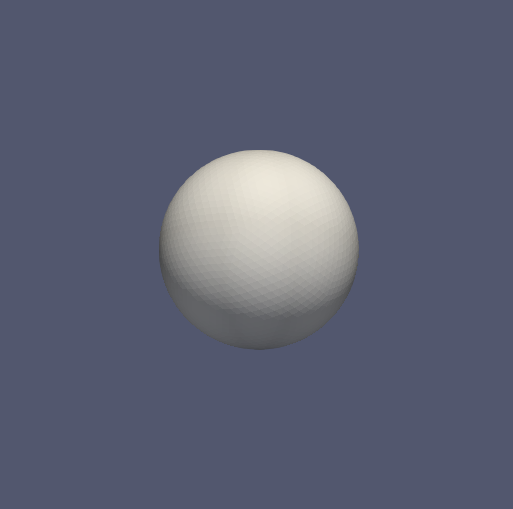} & \includegraphics[width=\fh,totalheight=\fh]{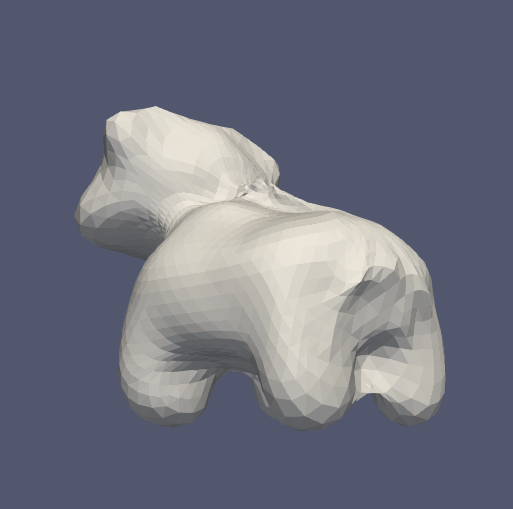} & \includegraphics[width=\fh,totalheight=\fh]{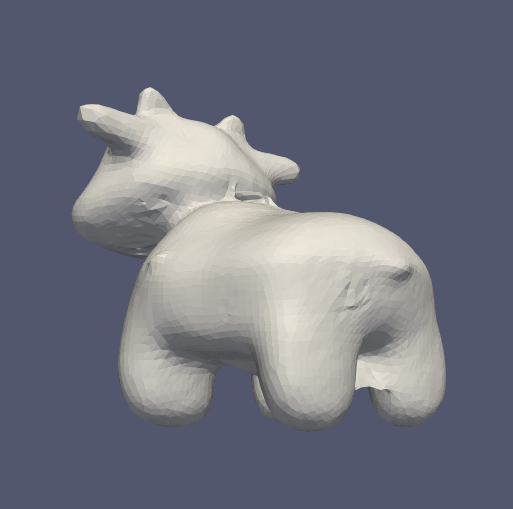} & \includegraphics[width=\fh,totalheight=\fh]{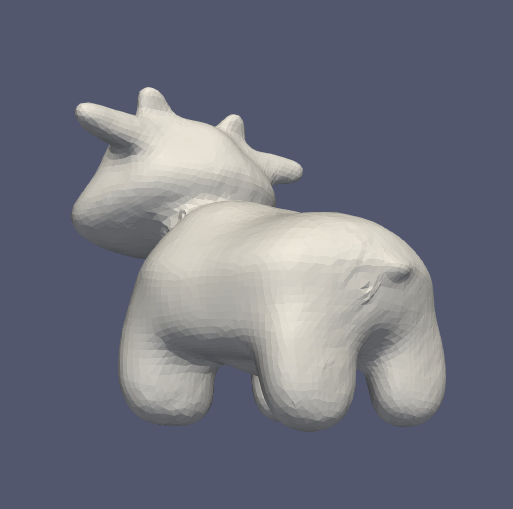} & \includegraphics[width=\fh,totalheight=\fh]{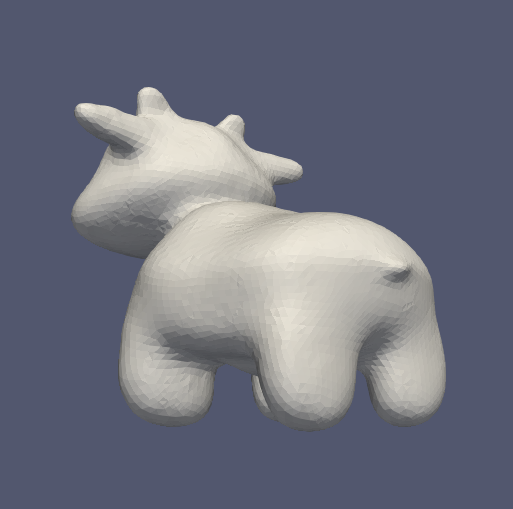} \\
\includegraphics[width=\fh,totalheight=\fh]{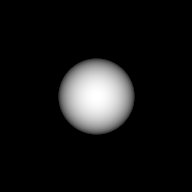} & \includegraphics[width=\fh,totalheight=\fh]{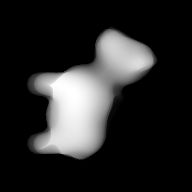} & \includegraphics[width=\fh,totalheight=\fh]{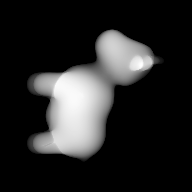} & \includegraphics[width=\fh,totalheight=\fh]{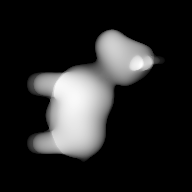} & \includegraphics[width=\fh,totalheight=\fh]{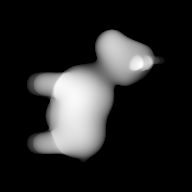} \\
\includegraphics[width=\fh,totalheight=\fh]{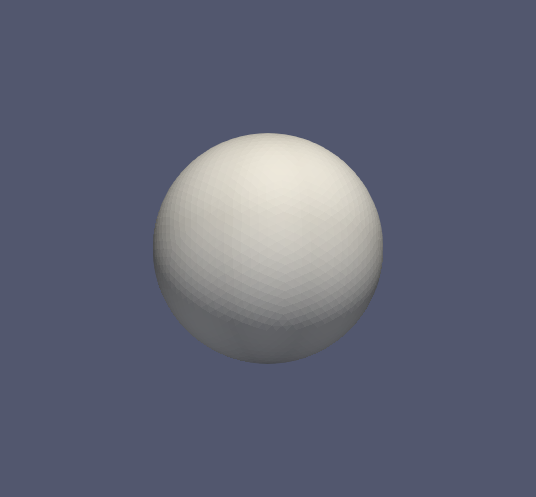} & \includegraphics[width=\fh,totalheight=\fh]{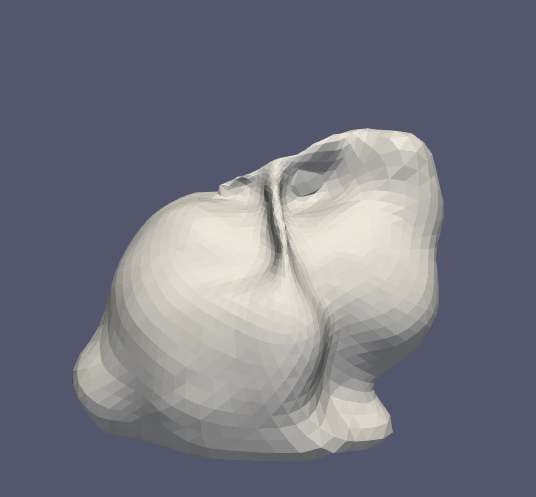} & \includegraphics[width=\fh,totalheight=\fh]{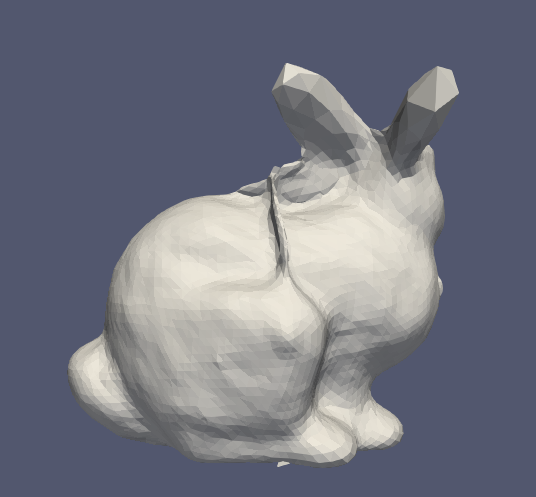} & \includegraphics[width=\fh,totalheight=\fh]{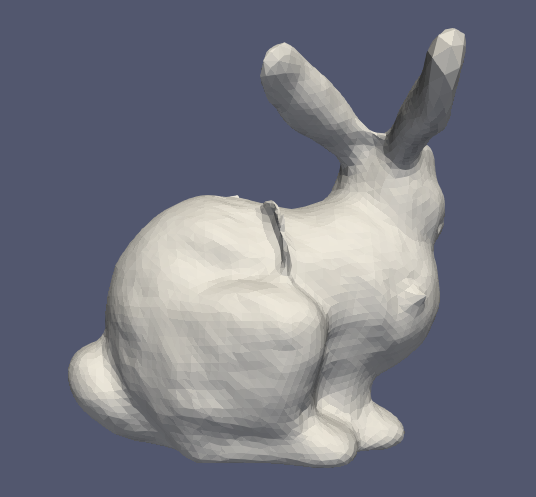} & \includegraphics[width=\fh,totalheight=\fh]{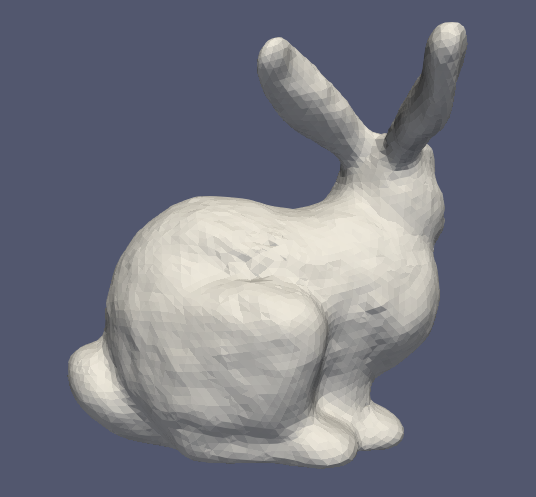} \\
\includegraphics[width=\fh,totalheight=\fh]{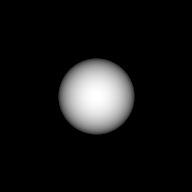} & \includegraphics[width=\fh,totalheight=\fh]{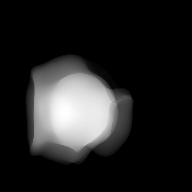} & \includegraphics[width=\fh,totalheight=\fh]{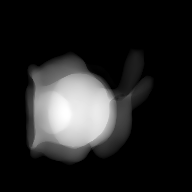} & \includegraphics[width=\fh,totalheight=\fh]{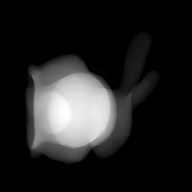} & \includegraphics[width=\fh,totalheight=\fh]{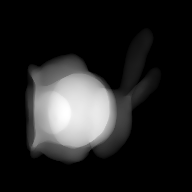} \\
\includegraphics[width=\fh,totalheight=\fh]{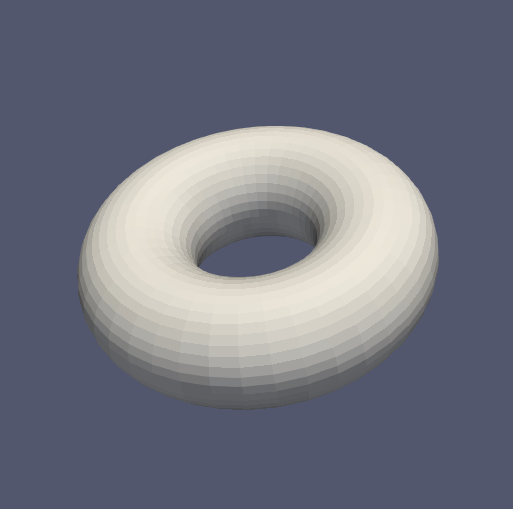} & \includegraphics[width=\fh,totalheight=\fh]{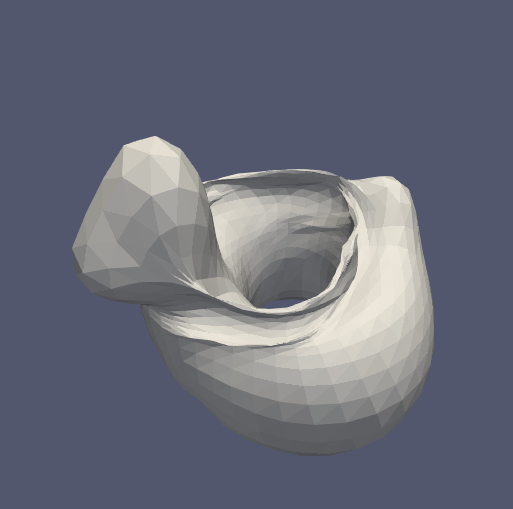} & \includegraphics[width=\fh,totalheight=\fh]{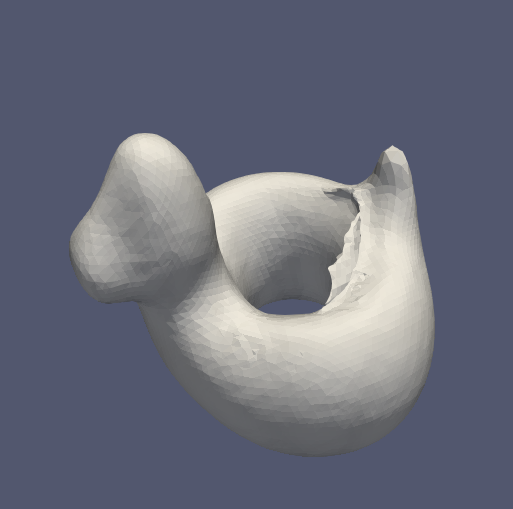} & \includegraphics[width=\fh,totalheight=\fh]{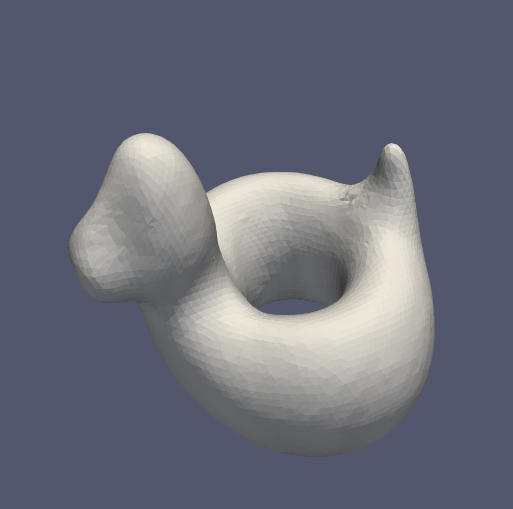} & \includegraphics[width=\fh,totalheight=\fh]{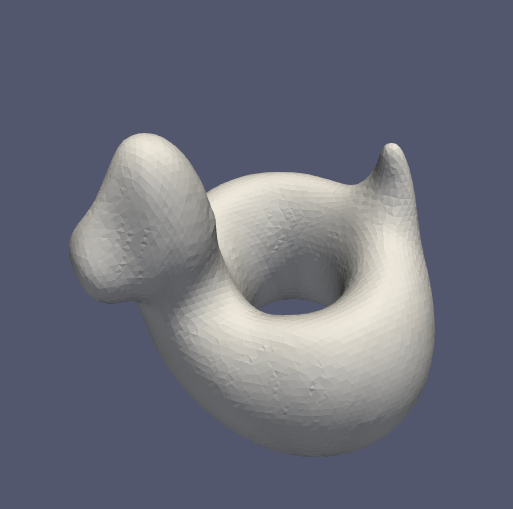} \\
\includegraphics[width=\fh,totalheight=\fh]{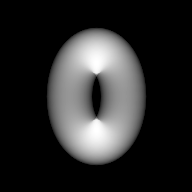} & \includegraphics[width=\fh,totalheight=\fh]{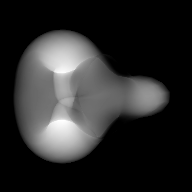} & \includegraphics[width=\fh,totalheight=\fh]{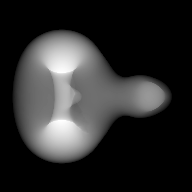} & \includegraphics[width=\fh,totalheight=\fh]{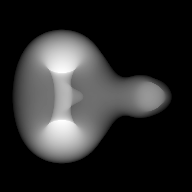} & \includegraphics[width=\fh,totalheight=\fh]{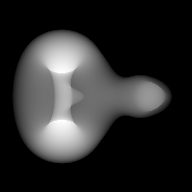} \\
init. & iter. 60 & iter. 120 & iter. 180 & iter. 360 \\
\end{tabular}
\caption{Deformation examples. The odd rows show the intermediate meshes during deformation and the even rows show the corresponding computed projections for one projection angle. We refine the mesh in finer resolution at iteration 60 and fix the mesh at iterations 120 and 180.}
\label{fig:deform}
\end{figure}

\section{Experiments and Results}

In this section, we present the experimental results of the proposed method on synthetic data of single objects. We also show the results on real data of some nano particles from electron tomography, which has limited range of angles.

\def\fh{55pt}
\def\fw{40pt}
\begin{figure*}[tb]
\begin{center}
\begin{tabular}{c@{ }c@{ }c@{ }c@{ }c@{ }c}
\parbox[b][\fh][c]{\fw}{GT} & 
\includegraphics[totalheight=\fh]{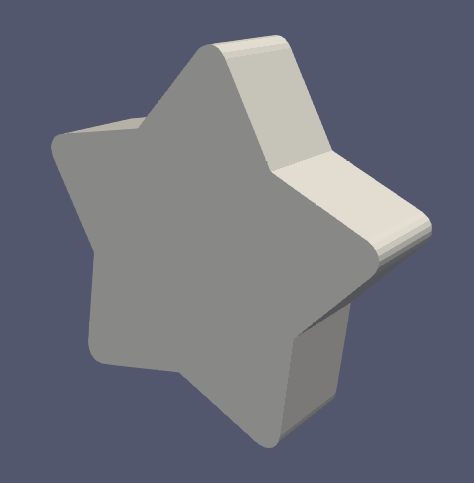} & \includegraphics[totalheight=\fh]{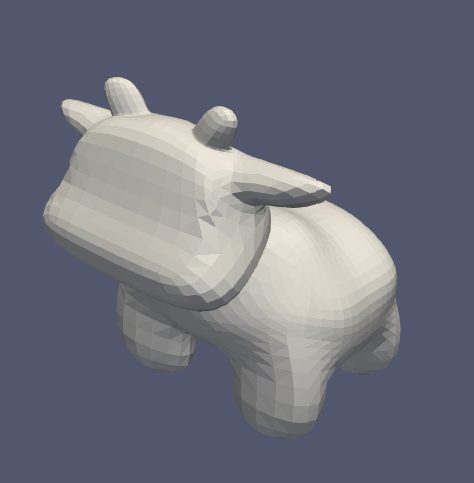} & \includegraphics[totalheight=\fh]{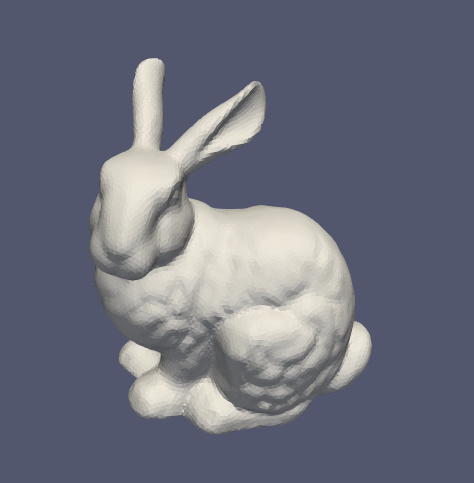} & \includegraphics[totalheight=\fh]{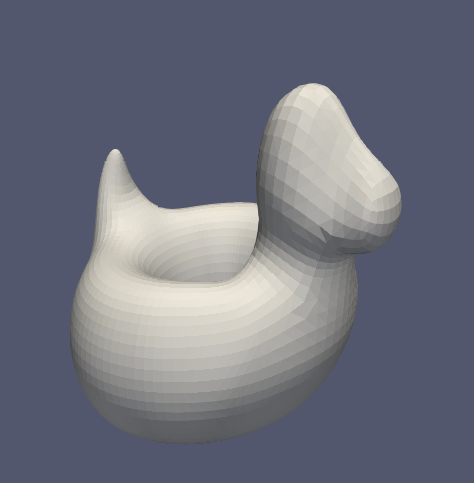} & \includegraphics[totalheight=\fh]{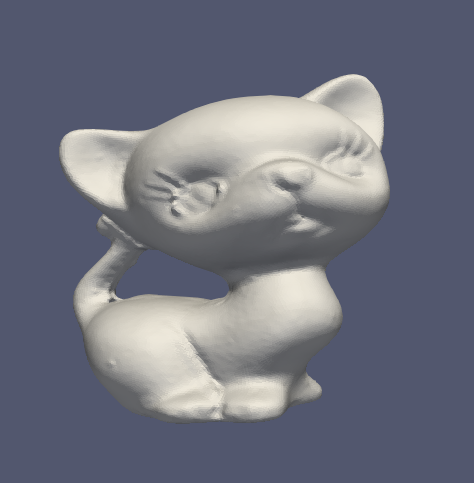} \\
\parbox[b][\fh][c]{\fw}{TV\\best} & 
\includegraphics[totalheight=\fh]{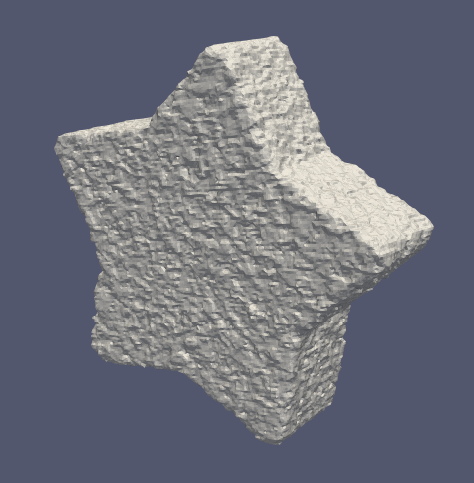} & \includegraphics[totalheight=\fh]{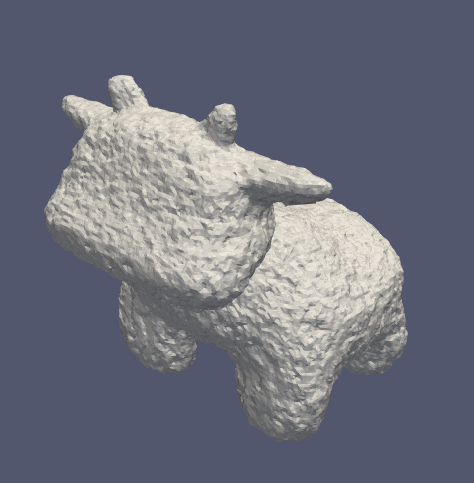} & \includegraphics[totalheight=\fh]{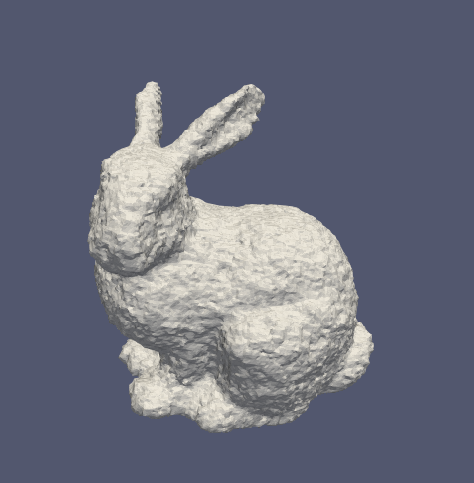} & \includegraphics[totalheight=\fh]{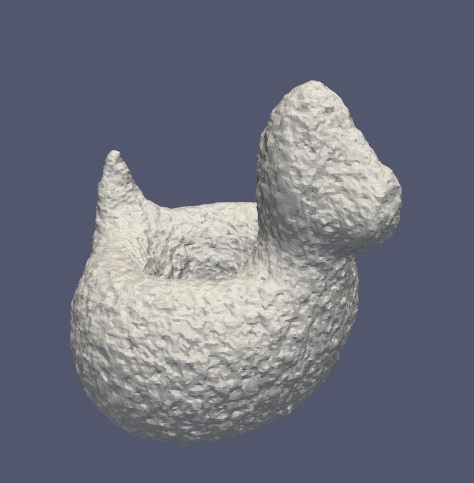} & \includegraphics[totalheight=\fh]{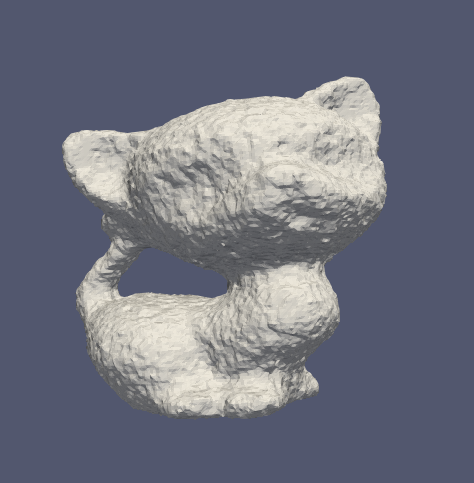} \\
\parbox[b][\fh][c]{\fw}{TV\\high} & 
\includegraphics[totalheight=\fh]{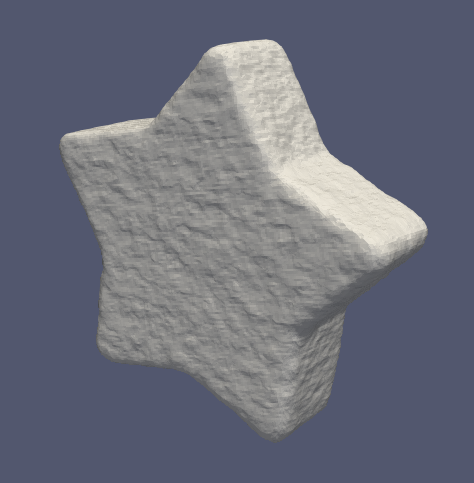} & \includegraphics[totalheight=\fh]{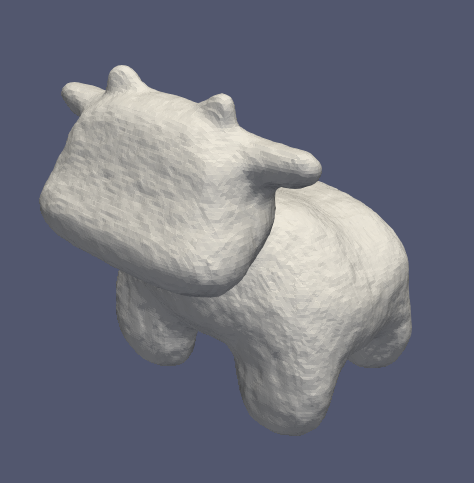} & \includegraphics[totalheight=\fh]{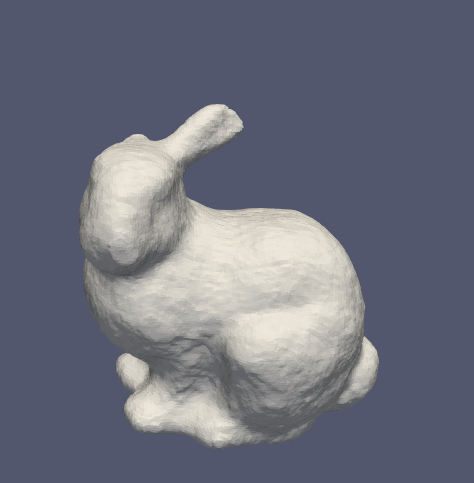} & \includegraphics[totalheight=\fh]{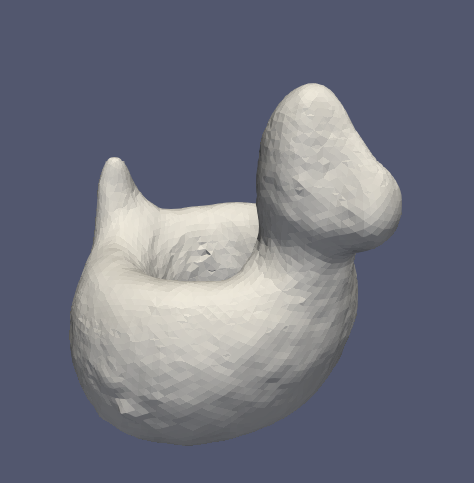} & \includegraphics[totalheight=\fh]{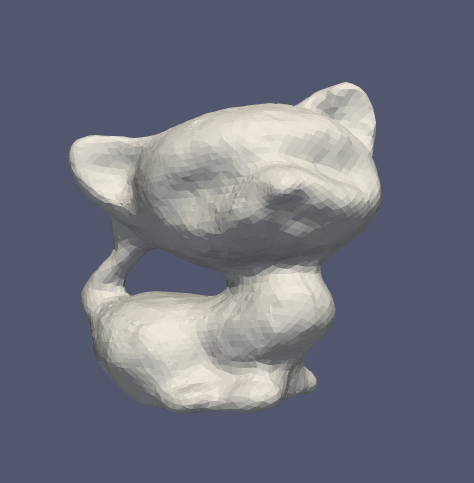} \\
\parbox[b][\fh][c]{\fw}{Proposed} & 
\includegraphics[totalheight=\fh]{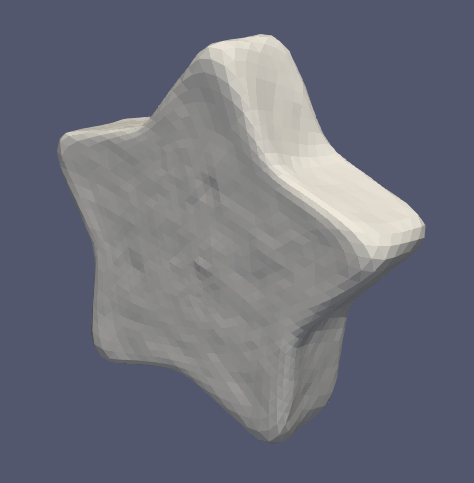} & \includegraphics[totalheight=\fh]{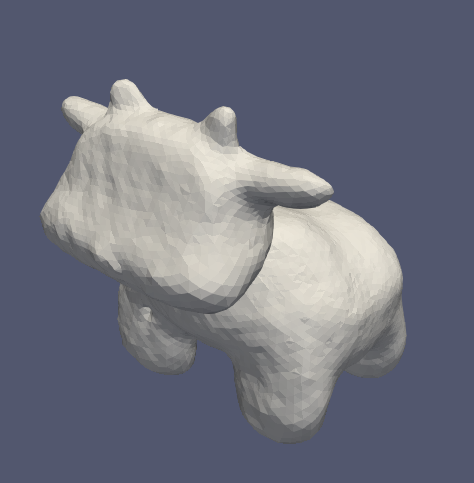} & \includegraphics[totalheight=\fh]{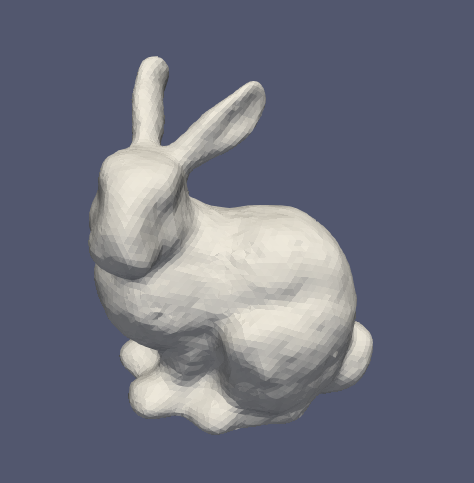} & \includegraphics[totalheight=\fh]{exp1/spinel/32} & \includegraphics[totalheight=\fh]{exp1/spinel/42} \\
\end{tabular}
\caption{Qualitative results of extracted meshes on noisy data with relative noise level 0.4. The top row shows the ground truth meshes. Rows 2 and 3 show the extracted isosurface from the results of TV reconstruction with the optimal regularization parameter (best) and a high regularization parameter (high), respectively. The last row shows our results.}
\label{fig:mesh}
\end{center}
\end{figure*}

\subsection{Shape reconstruction of a set of single objects}

\paragraph{Datasets}
This experiment is designed to test our shape estimation method on noisy simulated data. We use 5 watertight meshes (closed surfaces without any holes): star, spot, bunny, bob, kitten, shown in the first row of Fig.~\ref{fig:mesh} and the attenuations of the objects are set to 1.
Generating projections of those meshes using our forward model may resemble to the so-called inverse crime~\citep{mueller2012linear}. To avoid it, we employ the Blender software to make projection data based on ray casting methods similar to~\citep{marinovszki2018efficient}. 
We use 3D parallel projection geometry with 30 projection angles and the detector of size $192 \times 192$ pixels. Some projection images without noise are shown in the first row of Fig.~\ref{fig:projection} for one projection angle.

\paragraph{Evaluation metrics}
We compare our result to two standard reconstruction methods: simultaneous iterative reconstruction technique (SIRT)~\citep{andersen1984simultaneous} and total variation (TV) based reconstruction~\citep{chambolle2011firstorder,sidky2012convex}.
These methods yield 3D images, whereas the proposed method produces surface mesh, making direct comparison of the main output challenging. For consistent comparison, we employ a residual-based metric: \textit{residual projection error}~\citep{roelandts2014reconstructed}, which measures the $L^2$ norm difference of data and computed projections. We impose the relative Gaussian noise on the original data and calculate the residual projection error between noise-free projections and the estimations of other methods and the proposed method. For SIRT and TV, the voxel size is set as $192 \times 192 \times 192$ and the algorithm parameters are chosen carefully.

\paragraph{Experimental details}
Our implementation relies on PyTorch~\citep{paszke2019pytorch} and uses Adam~\citep{kingma2015adam} as an optimizer. The proposed forward projector is implemented as a module in PyTorch. As for the regularization parameters, we fix $\alpha=10$, $\gamma=0.01$ and iterate 500 times. The step size $\tau$ (learning rate in PyTorch) is set to 0.01 and reduced by half at 400 iteration. This reduction step is not really necessary but can yield a more stable result. We observe that 500 iterations are needed for capturing fine details of the complicated objects such as the \textit{bunny} data. During the experiment, we only vary the edge length parameter $\beta$ among the values of 1, 2, 4, 8, 16 and 32 and the optimal parameter would depend on the data. Finding the optimal regularization parameter is itself a research topic and not straightforward also in regularization-based image reconstruction methods such as TV. As for the initialization, the proposed method begins from an icosphere for genus-0 objects (\textit{star, spot, bunny}), and from a torus for genus-1 objects (\textit{bob, kitten}). Except for \textit{star} data, we refine the mesh by~\citep{huang2018robust} at iteration 60 and improve the mesh quality 3 times by a lightweight repair method~\citep{attene2010lightweight} at iteration 60, 120, 180. These refine and repair steps help remove some artifacts and lead to fast convergence. As also observed in \citep{vidal2009simulation}, some artifact pixels can appear (e.g., when the ray hits an odd number of times). When a large deformation happens, we may observe some artifact pixels, which we exclude in the objective function. However, in the end, we only observe around 2 artifact pixels. In Fig.~\ref{fig:deform}, we show the deformation of the estimated meshes with the corresponding computed projections. 
%

\def\fh{145pt}
\begin{figure}[!h]
\begin{center}
\includegraphics[totalheight=\fh]{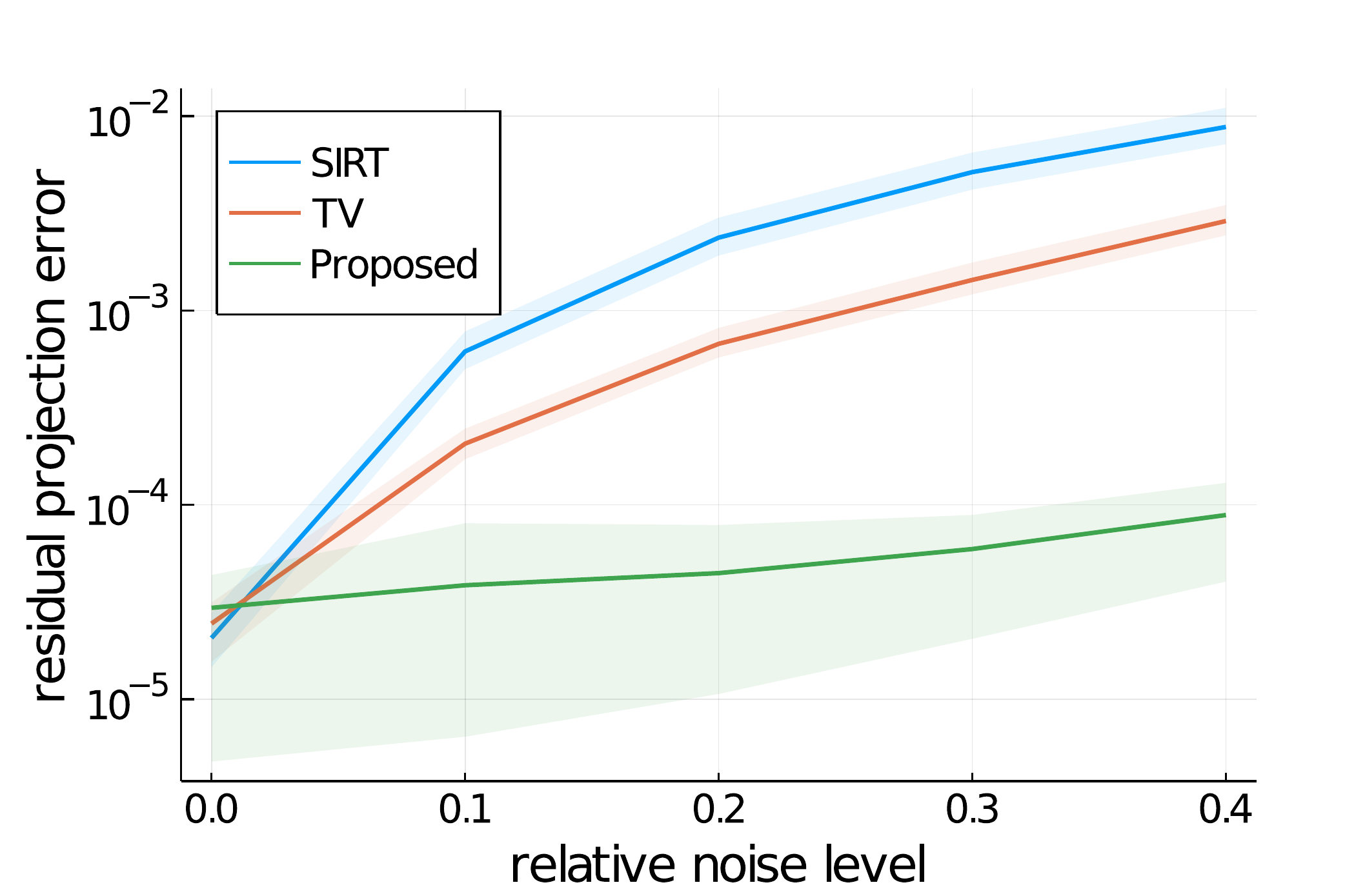}
\caption{Quantitative results with varying relative noise level over 5 datasets. The \textit{residual projection error} represents the error between the noise-free projections and the estimated projections. The error bars denote the average values with the maximum and minimum value and the y-axis is in logarithmic scale. }
\label{fig:graph}
\end{center}
\end{figure}

\paragraph{Robustness to noise}
The experiments show that the proposed method is robust to noise. Fig.~\ref{fig:projection} shows some computed projection images of the reconstruction results from noisy data with relative noise level 0.4 achieved using SIRT, TV reconstruction and the proposed method. Since SIRT has no regularization, it fits closely to the highly noisy data. The results of TV are relatively smooth, but sharp transitions appear blurred. On the other hand, the proposed method yields projections similar to noise-free data. Fig.~\ref{fig:mesh} shows the final mesh results, where the proposed method yields qualitatively better results than TV -- we omit highly noisy SIRT results.
In Fig.~\ref{fig:graph}, we provide the quantitative results of residual projection error with respect to relative noise levels. Without noise, SIRT gives the superior result as it fits to data without regularization. However, as noise increases, the results of SIRT and TV are shown to be poorer. 

\paragraph{Effect of parameters} 

Here, we investigate the effects of the parameters. In this experiment, we use data with a fixed relative noise level of 0.4 and use \textit{residual projection error} to measure performance. Fig.~\ref{fig:effect1} shows the effect of initializing the mesh by varying the radius of the icosphere used for initialization and then iterating 500 times. The final result is mostly not affected by the initial radius size thanks to the refinement steps during the optimization. We choose the $\textit{bunny}$ data model to generate the results shown in Fig.~\ref{fig:effect1}, because of its complicated shape.

In Fig.~\ref{fig:effect2} we show the effects of the regularization parameters $\alpha$, $\beta$ and $\gamma$, while keeping other parameters fixed. Here we use all data models and the residual projection error for the \textit{star} is large due to the coarse resolution of the mesh. We observe that if $\beta < 1.0$ or $\gamma > 0.1$, the final meshes might have some artifacts. As mentioned before, we use the default settings ($\alpha=10$ and $\gamma=0.01$), which in most experiments give a stable result. With these fixed values, we observe that it is enough to only vary the edge length parameter, depending on the desired degree of smoothness.

\def\fh{145pt}
\begin{figure}[!h]
\begin{center}
\includegraphics[totalheight=\fh]{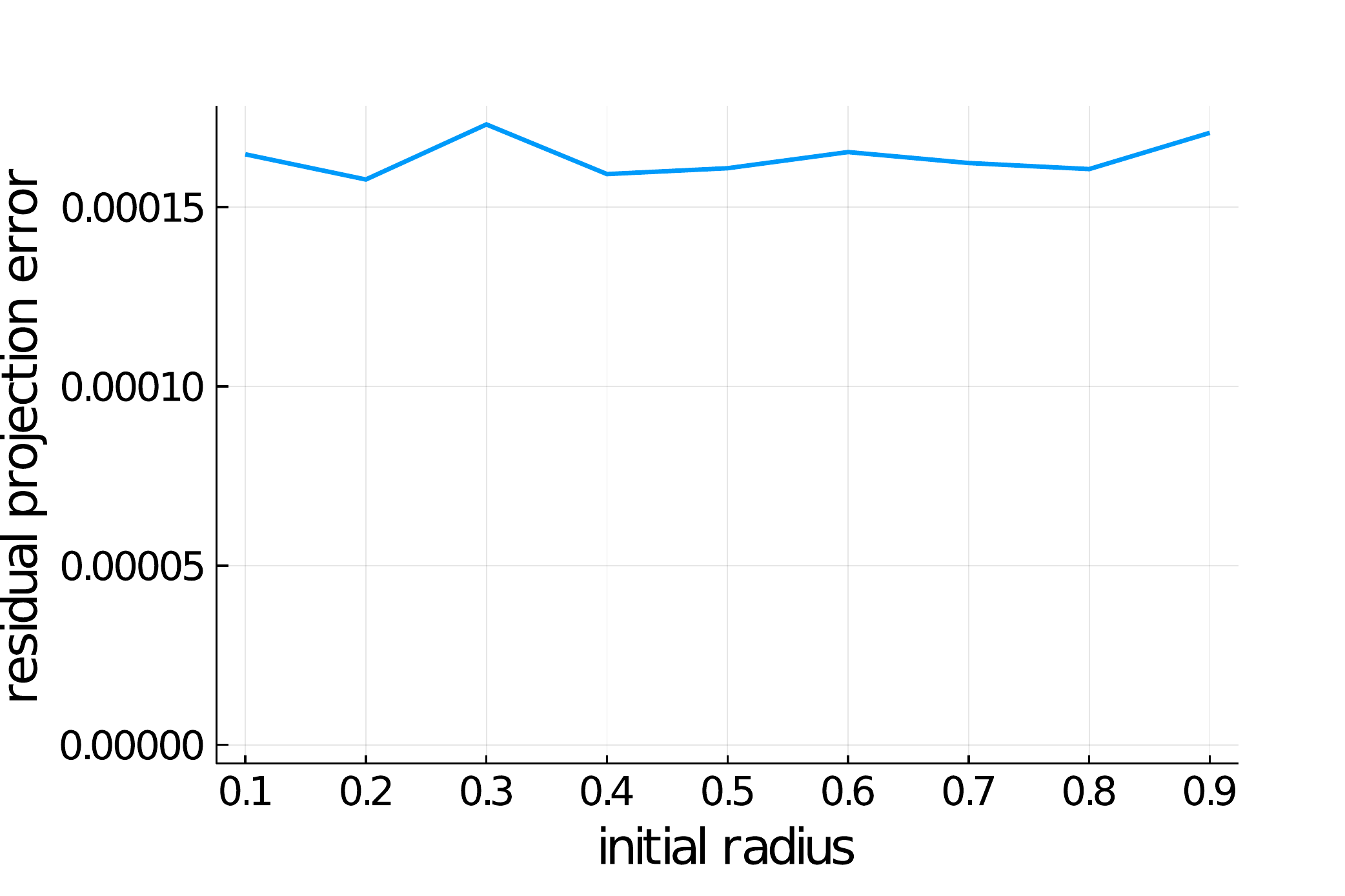}
\caption{Effect of initializing the mesh with spheres of varying diameter. The $x$-axis represents the radius of the initialized sphere. The radius is determined relatively to an object space that is normalized to $(-1, 1)^3$. The experiments are based on the \textit{bunny} data model.}
\label{fig:effect1}
\end{center}
\end{figure}

\def\fh{140pt}
\begin{figure}[!h]
\begin{center}
\begin{tabular}{c}
\includegraphics[totalheight=\fh]{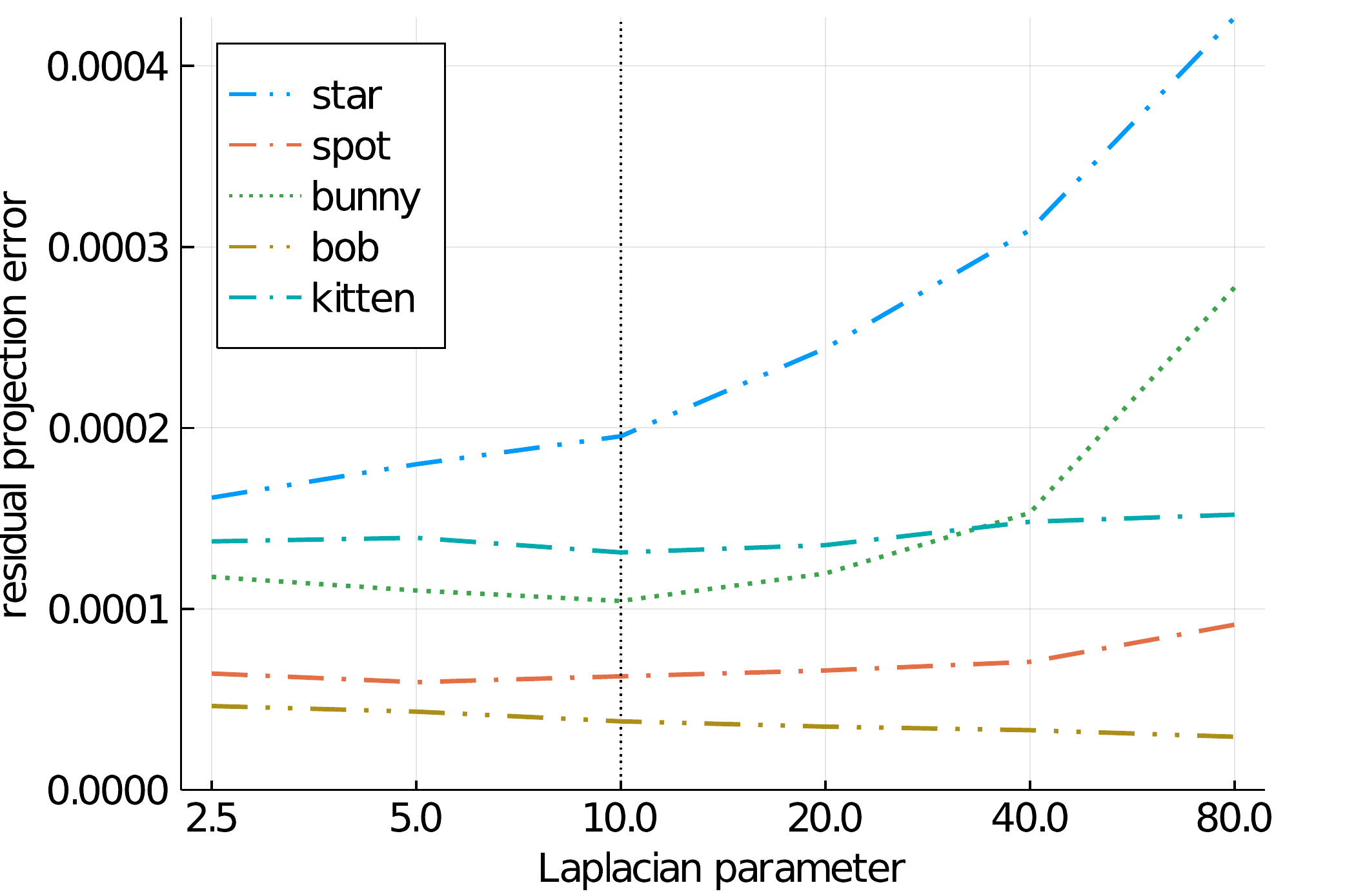}  \\
(a) Effect of $\alpha$ \\
\includegraphics[totalheight=\fh]{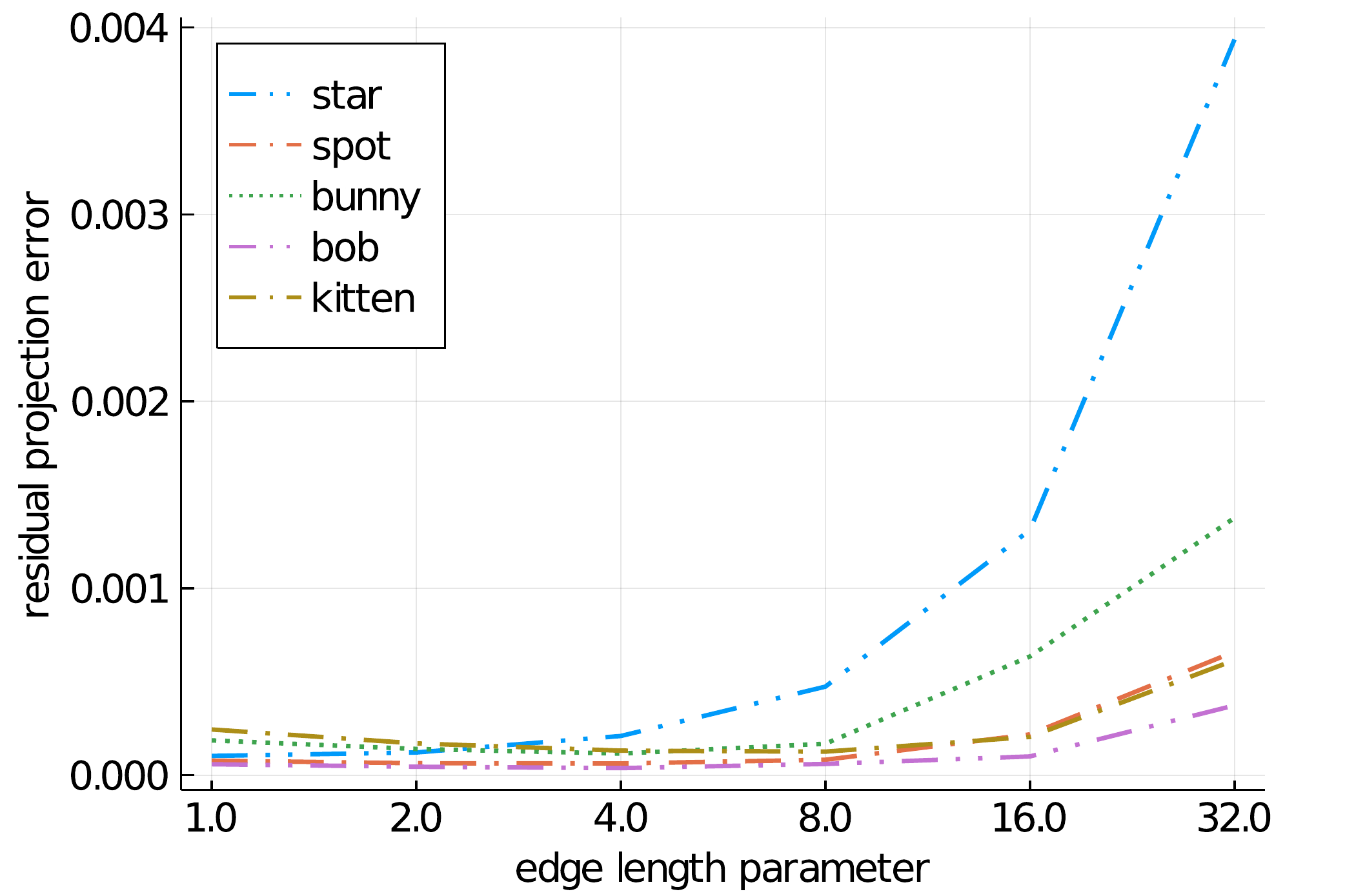}  \\
(b) Effect of $\beta$\\
\includegraphics[totalheight=\fh]{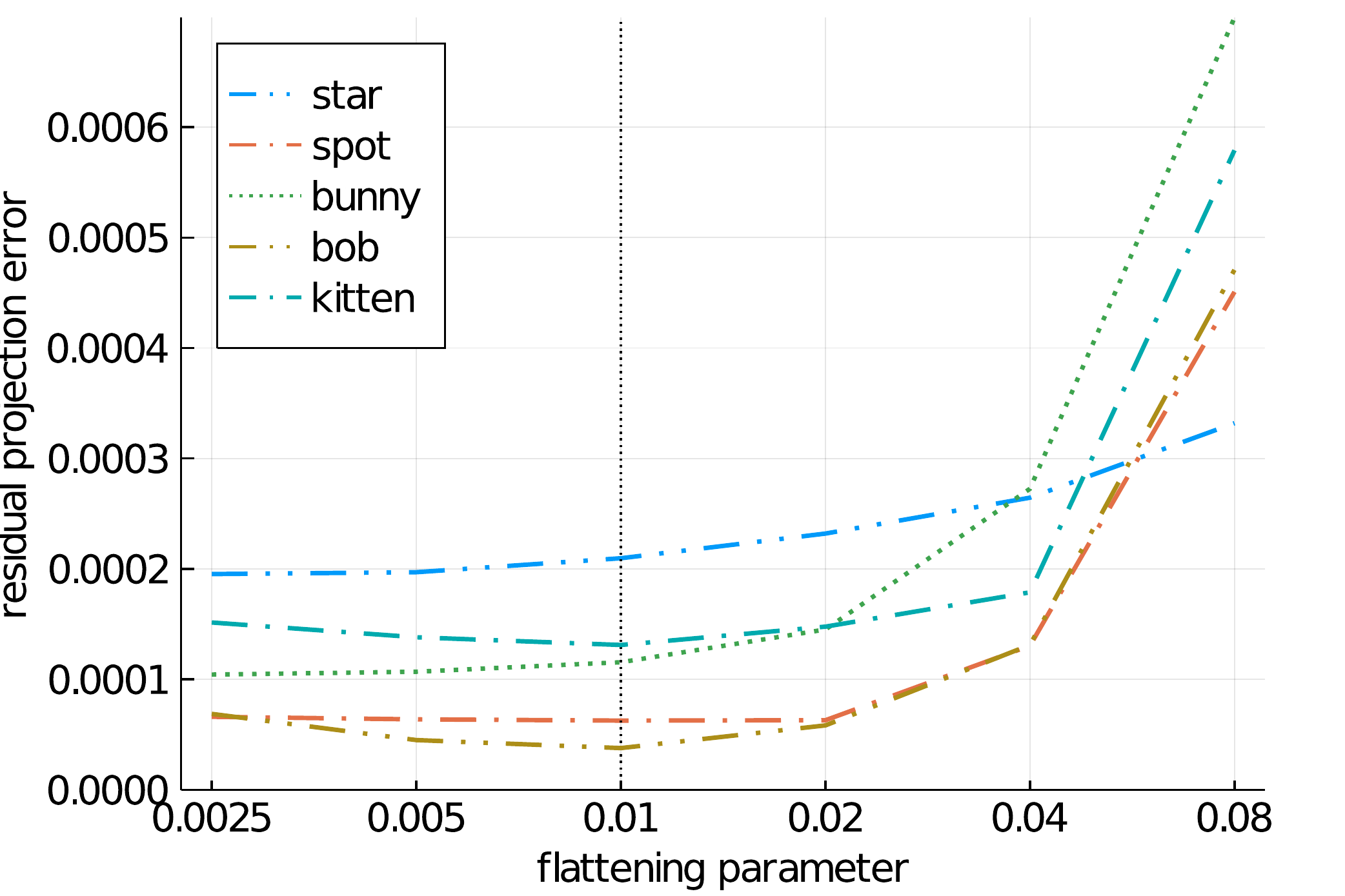}  \\
(c) Effect of  $\gamma$ \\
\end{tabular}
\caption{Effect of the regularization parameters: (a) the Laplacian regularization parameter $\alpha$ given $\beta=4$, $\gamma=0.01$, (b) the edge length parameter $\beta$ given $\alpha=10$, $\gamma=0.01$, and (c) the flattening parameter $\gamma$ given $\alpha=10$, $\beta=4$.
The vertical bars in (a) and (c) correspond to the values we keep fixed, when our method is compared with other methods.}
\label{fig:effect2}
\end{center}
\end{figure}

\paragraph{Computational cost}
Table~\ref{tab:statistics} shows the size of the initial and final mesh and the corresponding running times. The mesh size is one of the major factors that contribute to computational time. For example, as we do not refine the mesh for \textit{star} object, its final mesh size and its computational time is lower than others. We do the experiment on a Ubuntu server with 256GB RAM and Titan X GPU. 

For a reference, the running times for the image-based methods are 38 seconds for SIRT and 152 seconds for TV which are implemented based on multithreading on 8 CPU cores, not GPU. For image-based methods, we use the fixed grid of size 192 $\times$ 192 $\times$ 192 for all the data.

\setlength{\tabcolsep}{4pt}
\begin{table}[!h]
\centering
\caption{Size of meshes and running times} 
\begin{tabular}{|l | c c c | c c|}
\hline
Data                & Star       & Spot    & Bunny   & Bob     & Kitten  \\ \hline
Initial mesh & \multicolumn{3}{c|}{Icosphere (1280 faces)} & \multicolumn{2}{c|}{Torus (3200 faces)} \\
Final mesh & no refine. & 18680   & 17908   & 19308   & 20904   \\
Run time & 244.1 (sec.) & 730.7 & 719.6 & 816.3 & 816.8 \\ \hline
\end{tabular}
\label{tab:statistics}
\end{table}

\subsection{Application to electron tomography}

The goal of this experiment is to estimate the shape of a bimetallic nanoparticle having a Au-core and a Ag-shell nano particles. We obtained 2 tilt series datasets of a nano triangular bipyramid and a nanocube using high-angle annular dark-field scanning transmission electron microscopy (HAADF-STEM). We hereby used a Thermo Fisher Tecnai Osiris electron microscope generated at 200kV. Each dataset contains 49 projection images acquired over $\pm72^\circ$ with a tilt increment of 3º and a frame time of 4 seconds. This small range of projection angles is typical for electron tomography and makes the reconstruction challenging. Fig.~\ref{fig:proj} shows three images for each of the two datasets. 

\def\fh{70pt}
\begin{figure}[!ht]
\begin{center}
\begin{tabular}{c@{ }c@{ }c}
\includegraphics[totalheight=\fh]{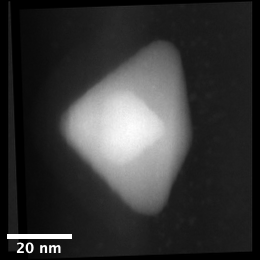} & 
\includegraphics[totalheight=\fh]{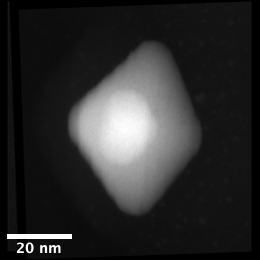} & \includegraphics[totalheight=\fh]{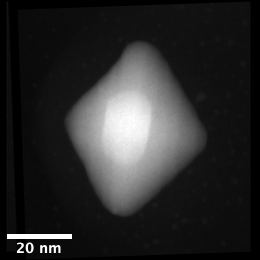}  \\
\includegraphics[totalheight=\fh]{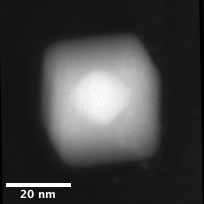} & 
\includegraphics[totalheight=\fh]{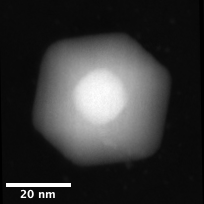} & \includegraphics[totalheight=\fh]{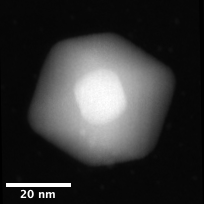}  \\
\end{tabular}
\caption{Projection images along 3 different tilt angles for a nano triangular bipyramid (top) and a nano cube (bottom).}
\label{fig:proj}
\end{center}
\end{figure}

To test the proposed method on the data, we initialize the meshes as two icospheres having total 2,560 faces. We use the collision detection method in~\citep{lauterbach2009fast}, to make sure that the core and the shell part do not collide. In this experiment, we set the number of iterations as 300, the step size $\tau$ as $0.005$ and observe no collisions when a proper regularization weight is used. As the unknown shapes are relatively simple, we impose only the Laplacian regularization term, by setting $\alpha=15$ for the nano bipyramid and $\alpha=5$ for the nanocube. The nanocube data is not much affected by this parameter $\alpha$, but for the nano bipyramid data, the parameter $\alpha$ should be greater than or equal to 10 to avoid the collision and obtain a high quality mesh when a small range of tilt angles is used. As for the initialization of two icospheres, the inner icosphere's size is set to half the size of the outer icosphere. This outer icosphere is initialized with a relative radius 0.4 (when the object space is normalized to $(-1,1)^3$), but we obtain a similar result with the radius range of 0.2 $\sim$ 0.8 for the nano cube and 0.2 $\sim$ 0.5 for the nano triangular data.

Fig.~\ref{fig:nanotri} and Fig.~\ref{fig:nanocube} show the 3D reconstruction for the nano triangular bypyramid and the nano cube data, respectively, by SIRT, TV, and the proposed method. To evaluate the effect of tilt angle range, we test two cases: an angular range of $\pm72^\circ$ and a subsampled tilt angles of $\pm18^\circ$. 
To compare our mesh result to the voxel-based methods, we extract meshes from the reconstruction images of SIRT and TV, by median filtering, thresholding, filling holes, and extracting the isosurfaces. For the highly-limited angle case with a tilt range of $\pm18^\circ$, we additionally apply Gaussian smoothing on the results of SIRT and TV before the thresholding step, to obtain the visually-appealing meshes.

For the tilt angle ranges of $\pm72^\circ$, the results of SIRT, TV, and the proposed method are in good comparison. However, when the range of tilt angles is reduced to $\pm18^\circ$, SIRT and TV yield a degenerated reconstruction, which also affects the quality of the final extracted surfaces. On the other hand, the proposed method is shown to be less affected by a tilt range. For the bypyramid in Fig.~\ref{fig:nanotri}, the volume of the Au-core decreased for the limited angle case, due to the high regularization effect. In Table~\ref{tab:volume}, we provide the estimated volumes of the Au-core and Ag-shell particles by the proposed method and the extracted isosurfaces from the images by SIRT and TV. 

\setlength{\tabcolsep}{6pt}
\begin{table}[h]
\caption{Estimated volumes in the unit of $10^3$ nm by SIRT reconstruction image followed by isosurface extraction, and the proposed method.  }
\centering
\begin{tabular}{| c | c | c c |  c c |}
\hline
\multirow{2}{*}{Tilt angles} & \multirow{2}{*}{Method}    & \multicolumn{2}{c|}{bipyramid} & \multicolumn{2}{c|}{nano cube} \\ \cline{3-6} 
  & & core & shell & core & shell\\ \hline
\multirow{3}{*}{-$72^\circ\sim72^\circ$}  & \begin{tabular}[c]{@{}l@{}}SIRT+Iso. \end{tabular} &   3.6  &    33.2   &2.4 &   29.8   \\
  & TV+Iso. &   3.7  &  33.5 &  2.5  & 29.8\\ 
  & Proposed &   4.2  &  36.7 &  2.3 & 31.2\\ \hline
\multirow{3}{*}{-$18^\circ\sim18^\circ$} & SIRT+Iso. & 1.3&31.5  & 1.2&  14.9    \\
  & TV+Iso. &  3.5   &  25.0&    1.3  & 11.2\\ 
  & Proposed &  2.7   &  37.7&    2.5  & 29.8\\ \hline
\end{tabular}
\label{tab:volume}
\end{table}

\def\fw{70pt}
\def\fh{95pt}
\begin{figure*}[h]
\begin{center}
\begin{tabular}{c@{ }c@{ }c@{ }c}
\parbox[b][\fh][c]{\fw}{SIRT+Isosurface\\\small{(-$72^\circ\sim72^\circ$)}} & 
\includegraphics[totalheight=\fh]{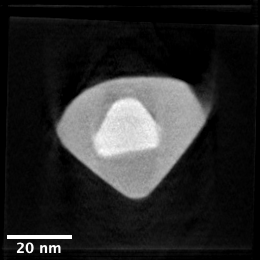} & \includegraphics[totalheight=\fh]{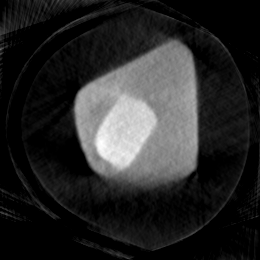} & \includegraphics[totalheight=\fh, trim={13cm 0cm 13cm 0cm}]{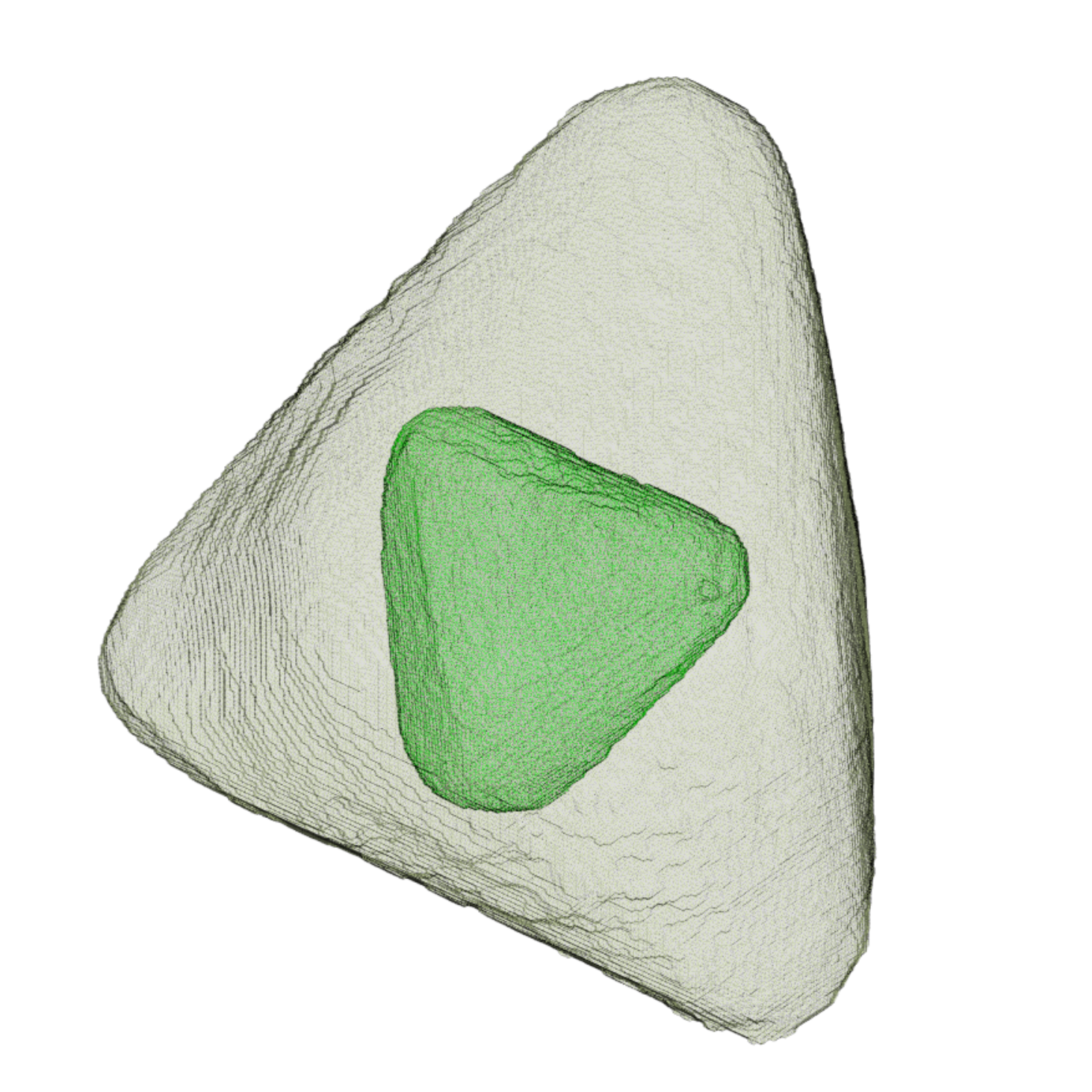} \\
\parbox[b][\fh][c]{\fw}{TV+Isosurface\\\small{(-$72^\circ\sim72^\circ$)}} & 
\includegraphics[totalheight=\fh]{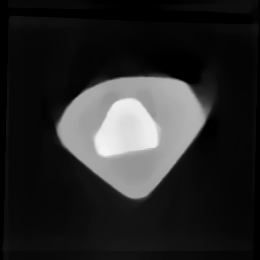} & \includegraphics[totalheight=\fh]{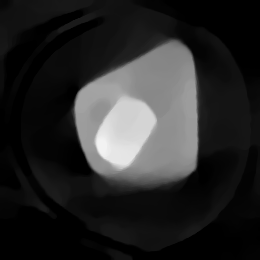} & \includegraphics[totalheight=\fh, trim={13cm 0cm 13cm 0cm}]{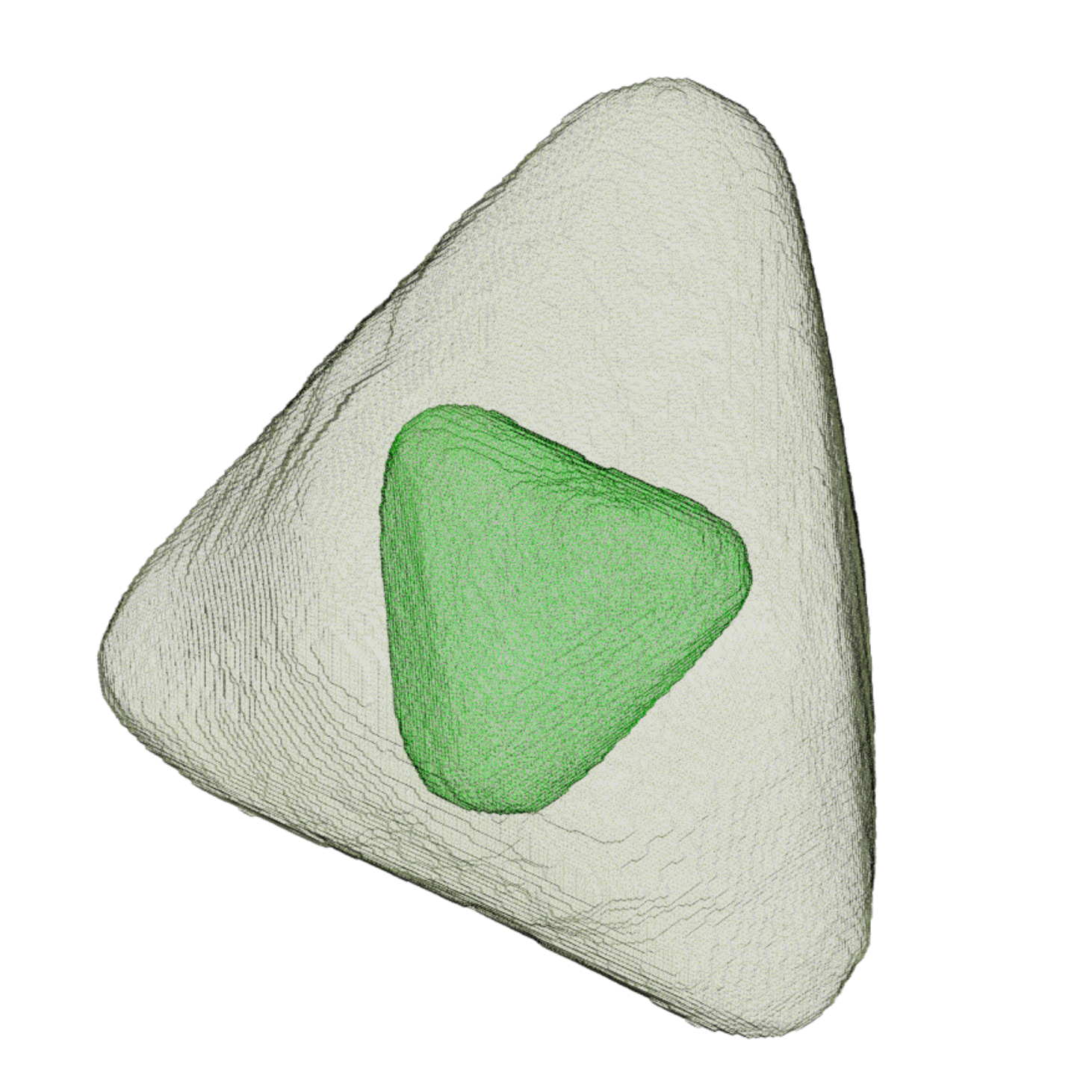} \\
\parbox[b][\fh][c]{\fw}{Proposed\\\small{(-$72^\circ\sim72^\circ$)}} & \includegraphics[totalheight=\fh]{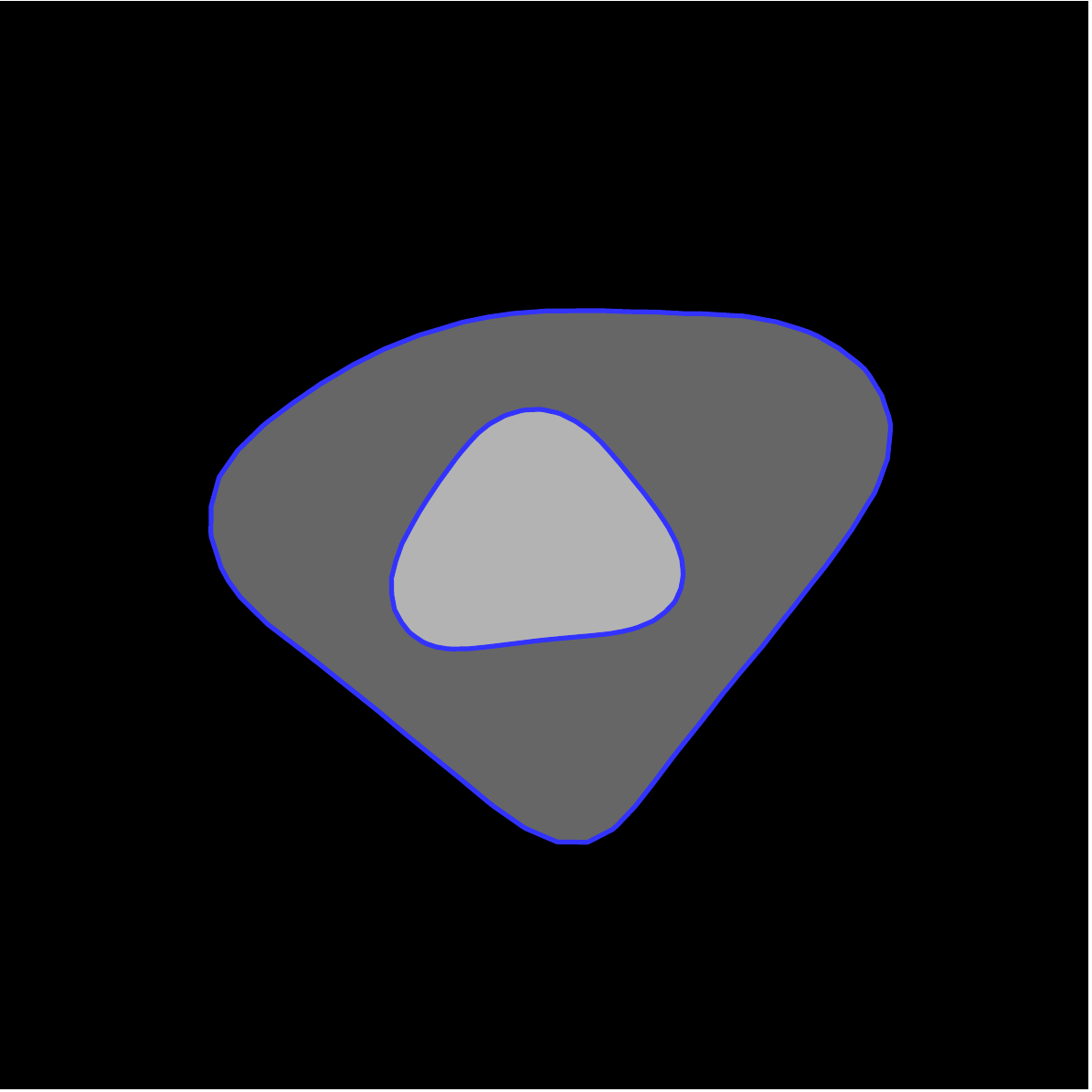} &\includegraphics[totalheight=\fh]{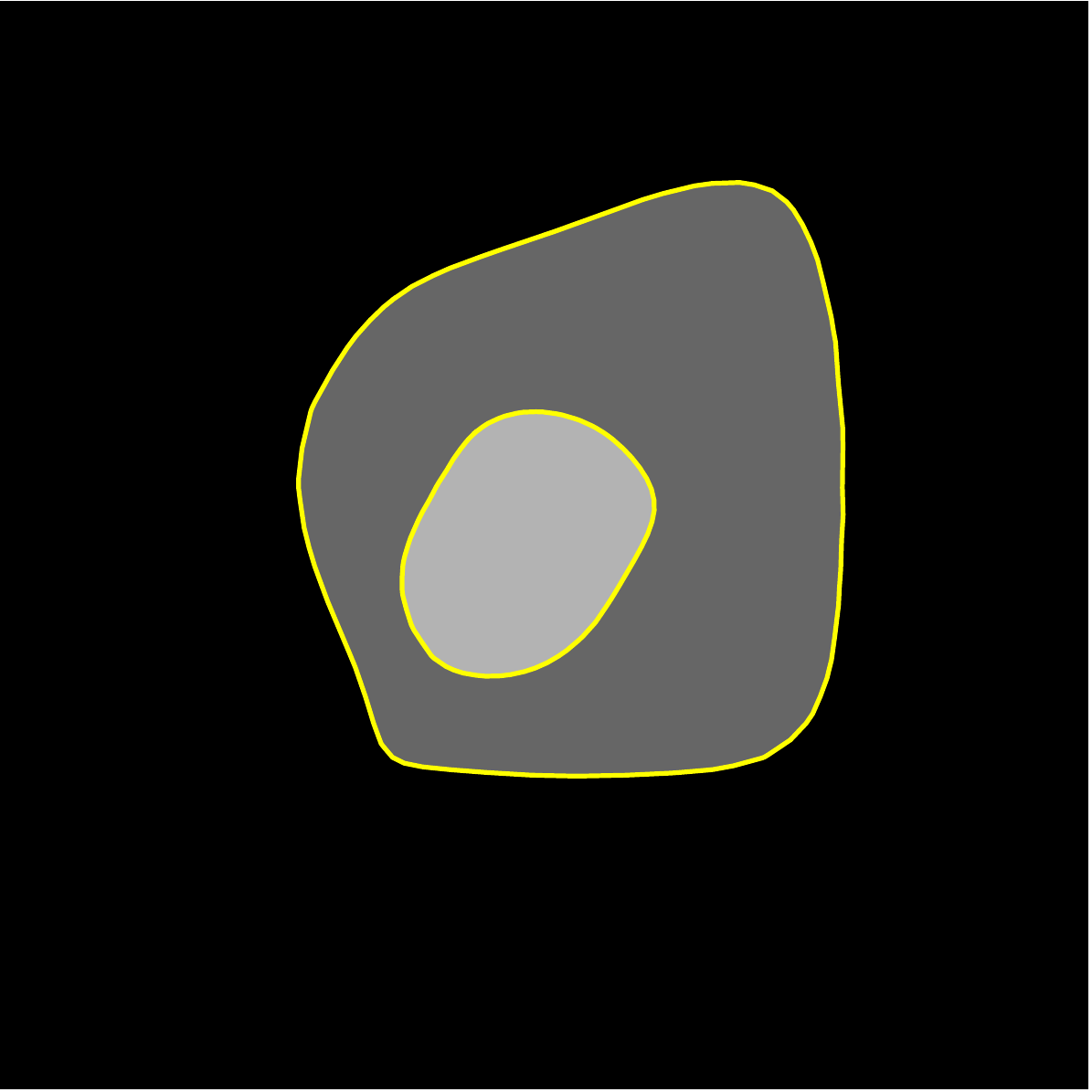} & \includegraphics[totalheight=\fh]{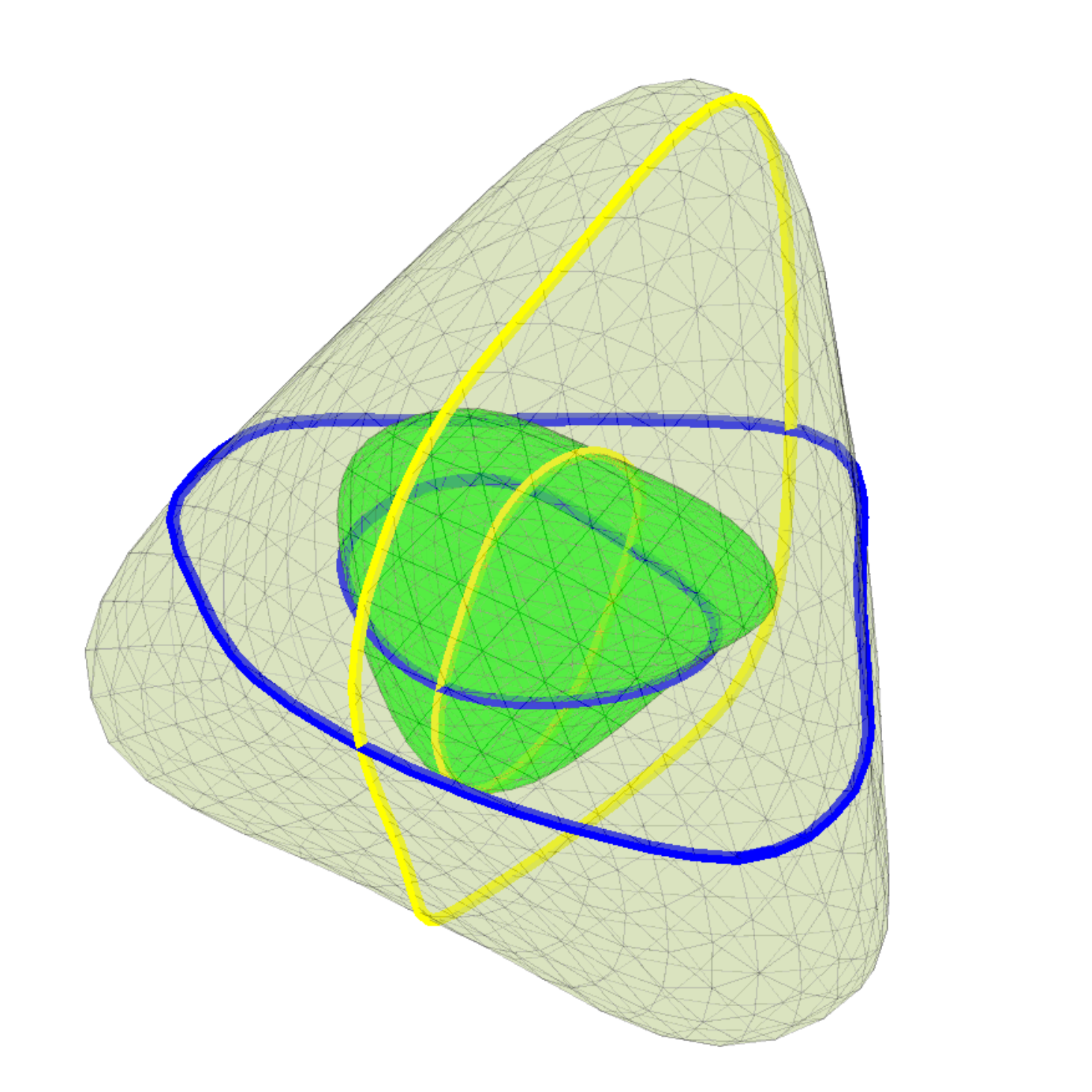} \\
\parbox[b][\fh][c]{\fw}{SIRT+Isosurface\\\small{(-$18^\circ\sim18^\circ$)}} & 
\includegraphics[totalheight=\fh]{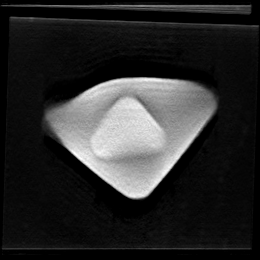} & \includegraphics[totalheight=\fh]{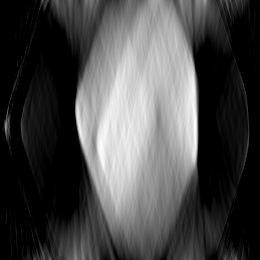} & \includegraphics[totalheight=\fh]{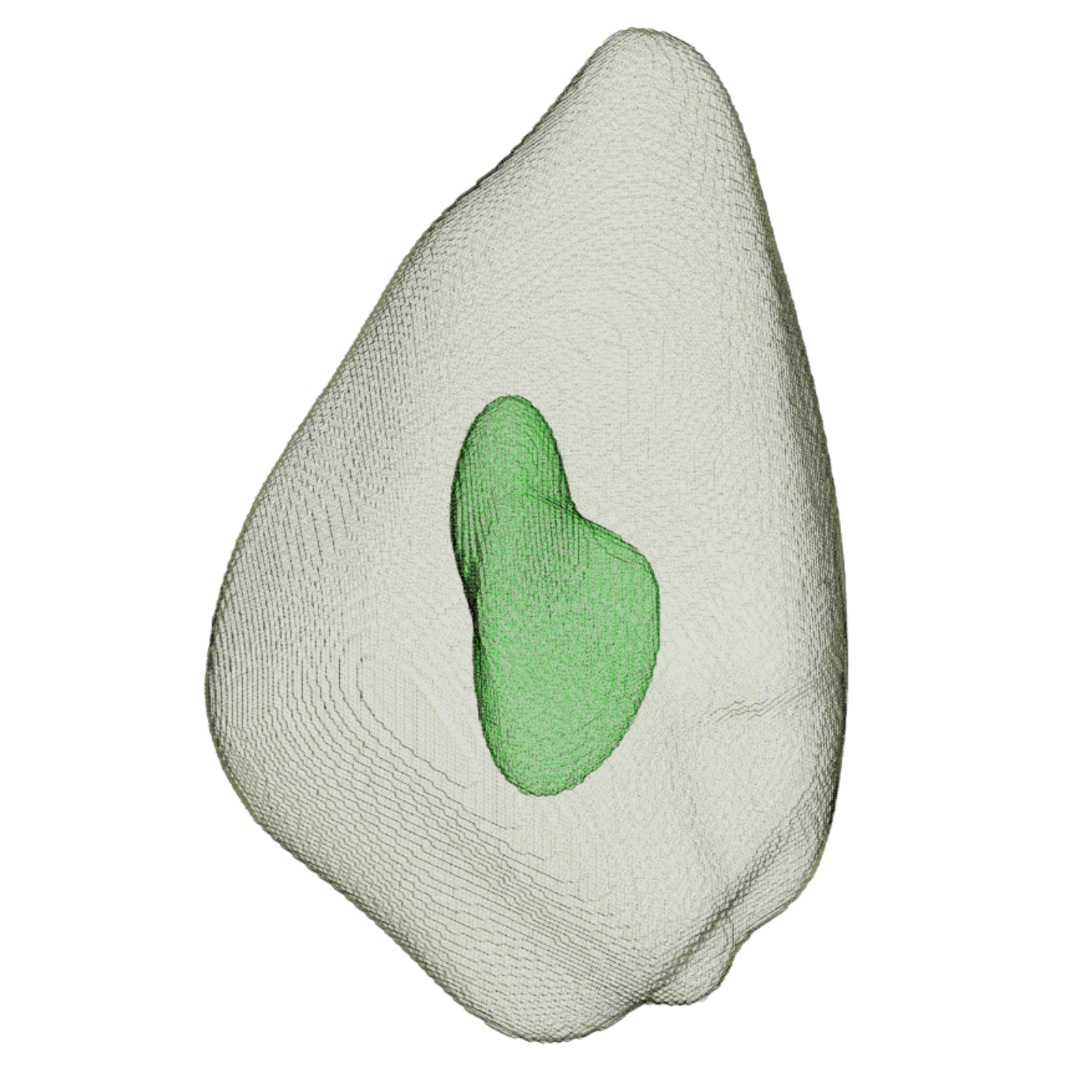} \\
\parbox[b][\fh][c]{\fw}{TV+Isosurface\\\small{(-$18^\circ\sim18^\circ$)}} & 
\includegraphics[totalheight=\fh]{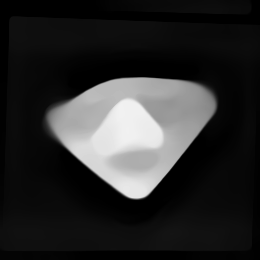} & \includegraphics[totalheight=\fh]{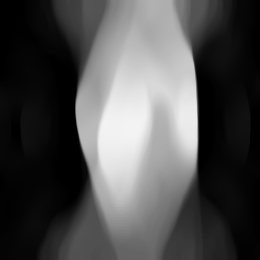} & \includegraphics[totalheight=\fh]{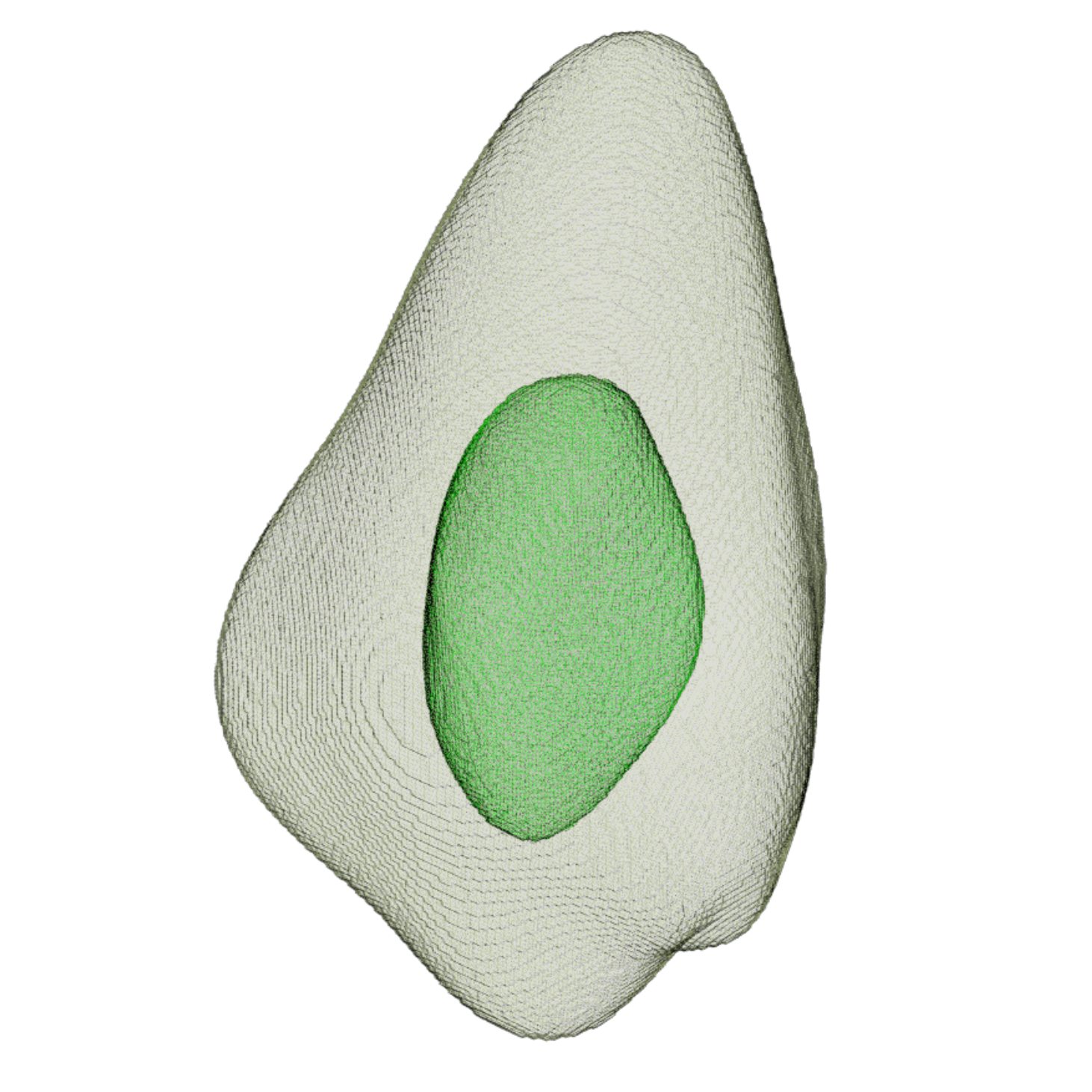} \\
\parbox[b][\fh][c]{\fw}{Proposed\\\small{(-$18^\circ\sim18^\circ$)}} & 
 \includegraphics[totalheight=\fh]{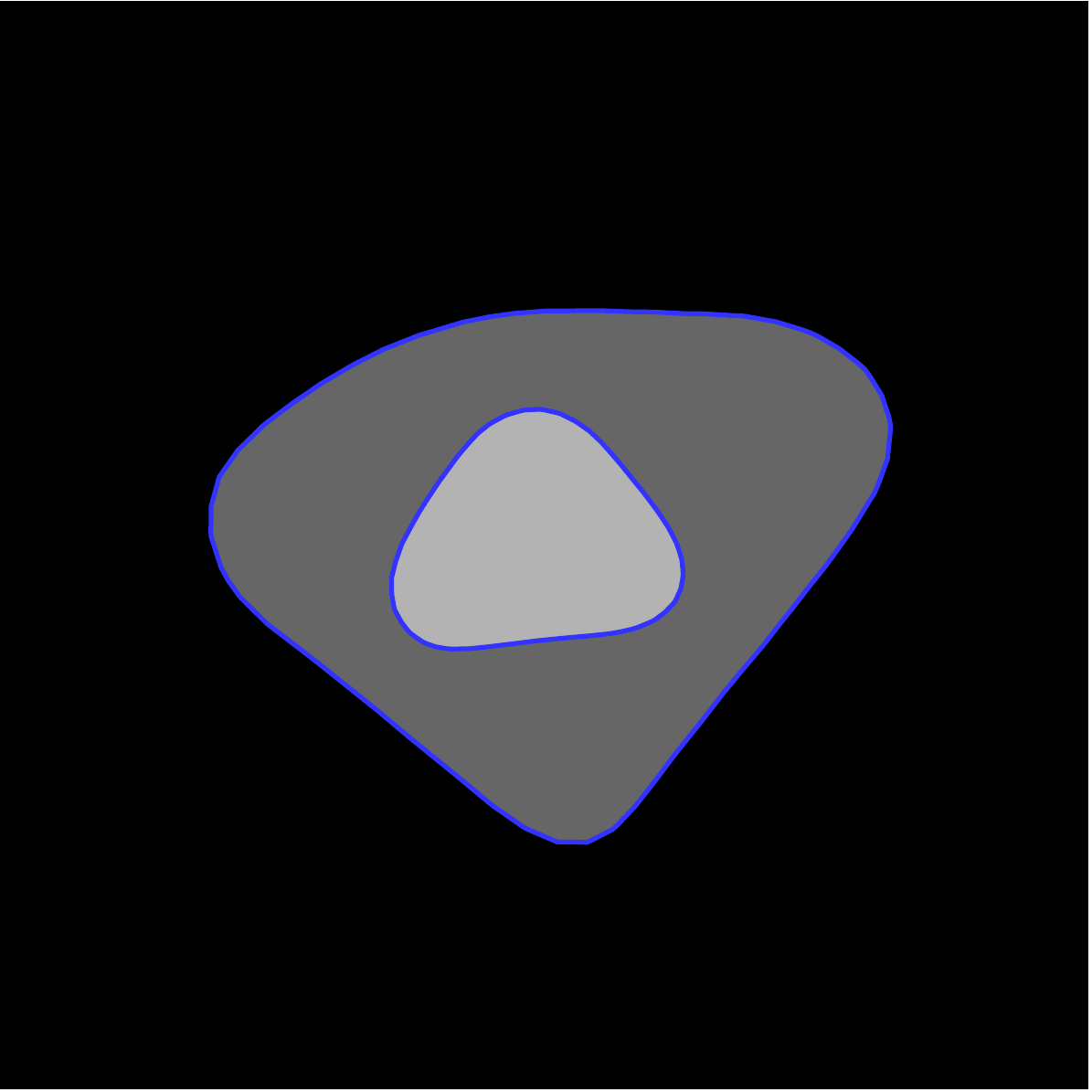} &\includegraphics[totalheight=\fh]{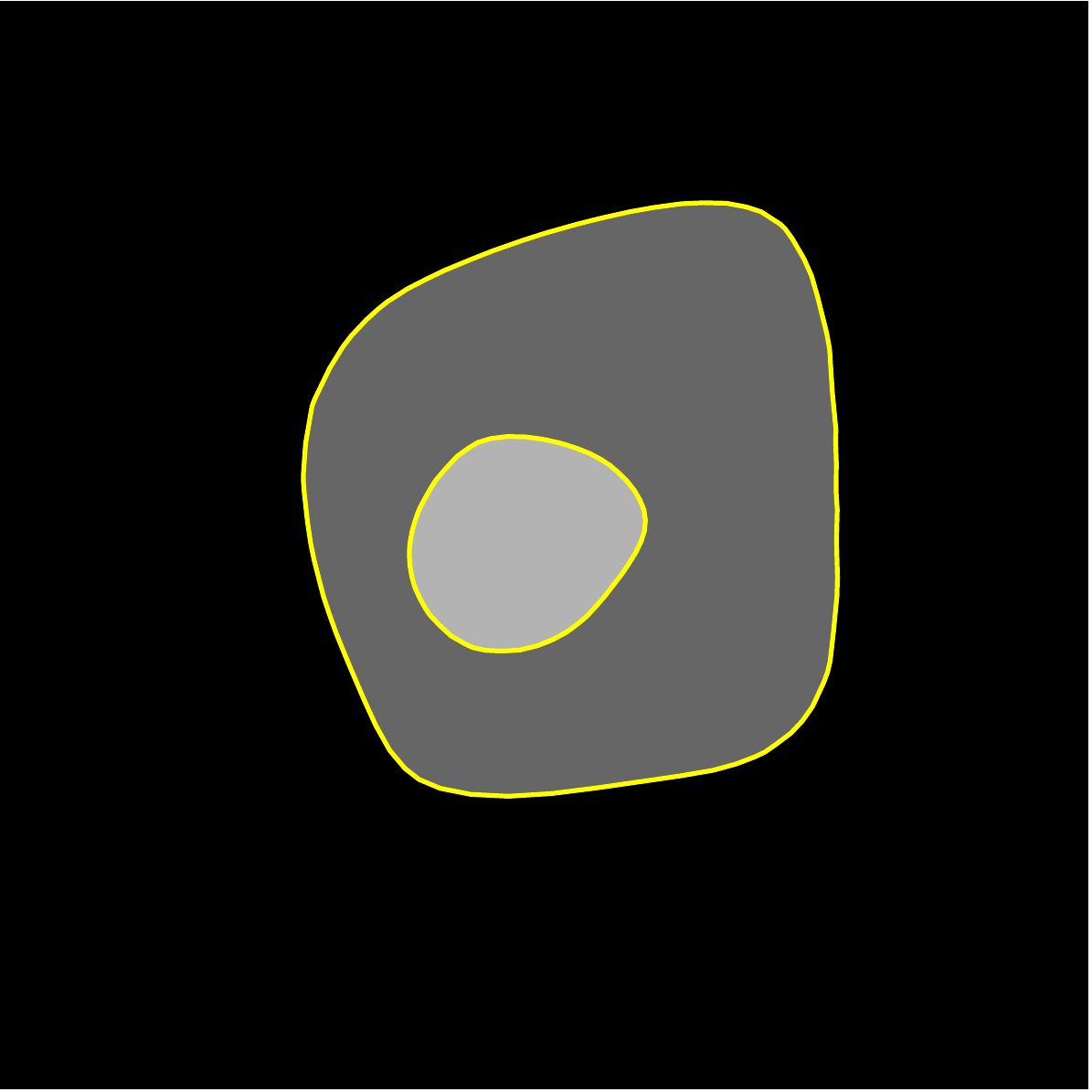}  & \includegraphics[totalheight=\fh]{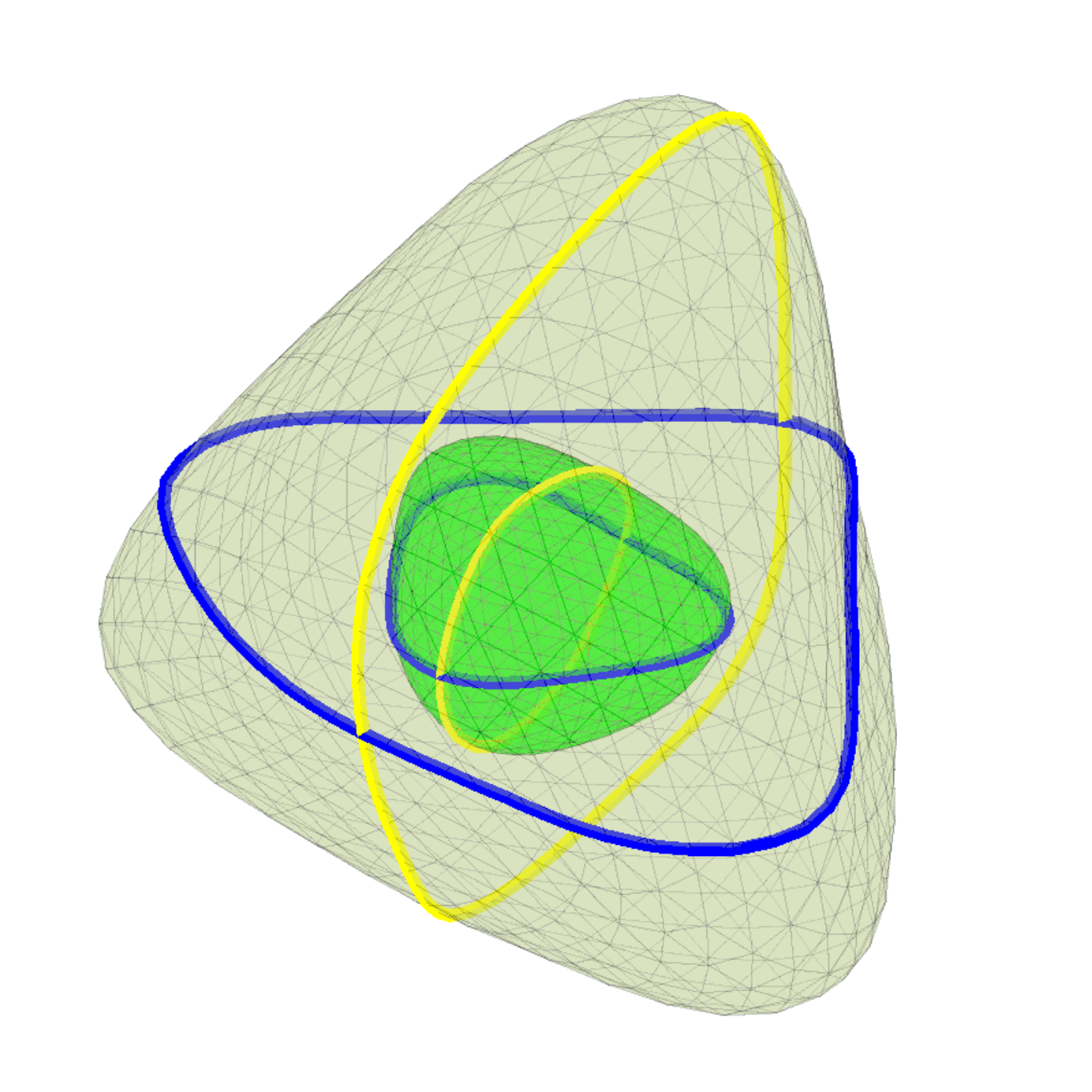}\\
\end{tabular}
\caption{Reconstruction results on nano triangular bypyramid data. The first and the 4th rows show the reconstruction results by SIRT and from projections with tilt angles from -72$^\circ$ to 72$^\circ$ and -18$^\circ$ to 18$^\circ$, respectively, where two central slices are visualized in the first and second column. The 2nd and the 5th rows show the reconstruction results by TV. The last column shows the extracted mesh from the reconstructed image by SIRT or TV with some post-processing procedures. The third and the last row show our direct shape estimation results, where two central slices are visualized in blue and yellow.}
\label{fig:nanotri}
\end{center}
\end{figure*}

\begin{figure*}[h]
\begin{center}
\begin{tabular}{c@{ }c@{ }c@{ }c}
\parbox[b][\fh][c]{\fw}{SIRT+Isosurface\\\small{(-$72^\circ\sim72^\circ$)}} & 
\includegraphics[totalheight=\fh]{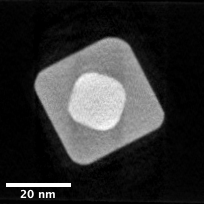} & \includegraphics[totalheight=\fh]{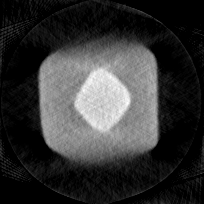} & \includegraphics[totalheight=\fh]{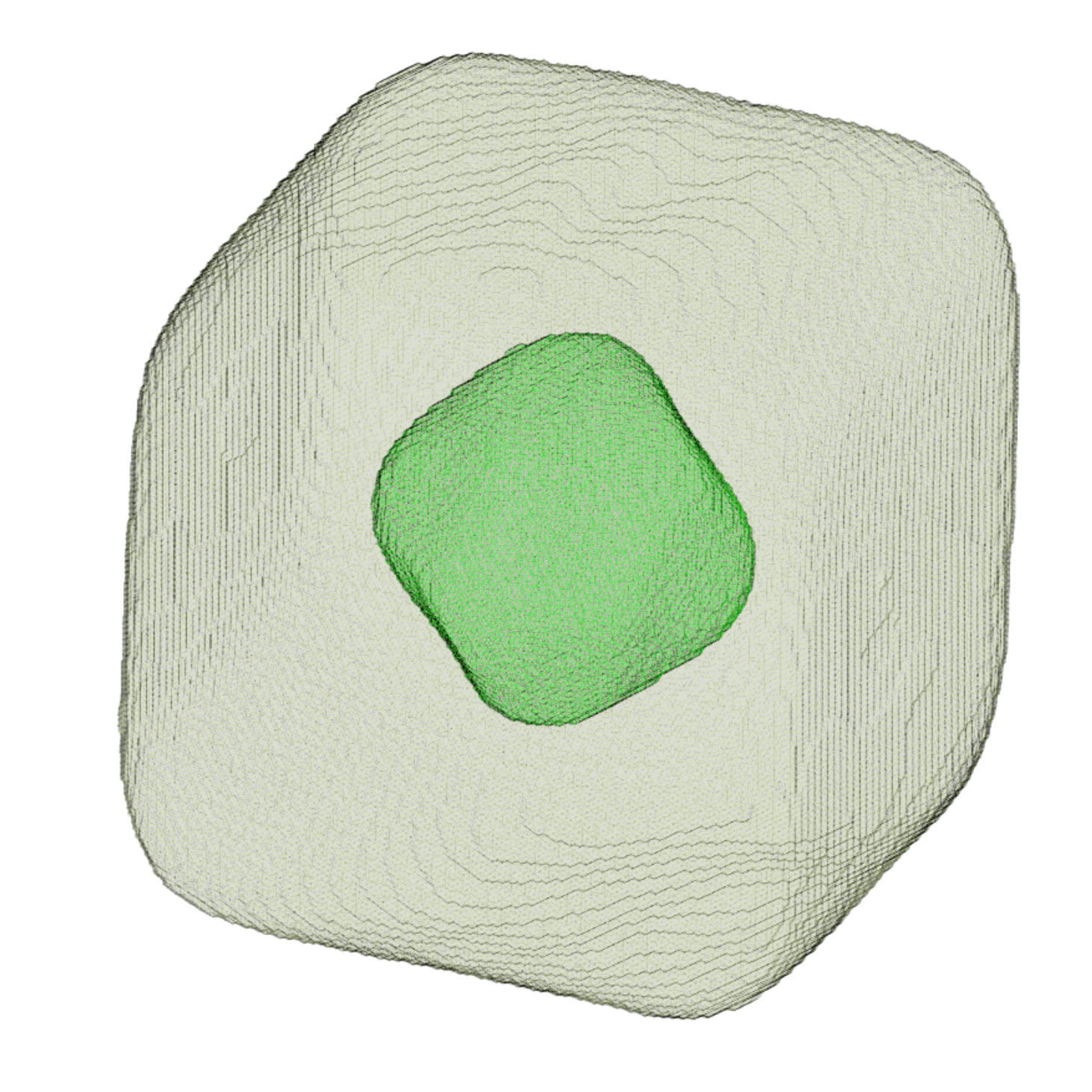} \\
\parbox[b][\fh][c]{\fw}{TV+Isosurface\\\small{(-$72^\circ\sim72^\circ$)}} & 
\includegraphics[totalheight=\fh]{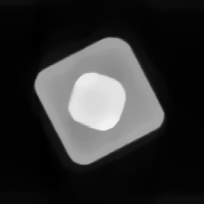} & \includegraphics[totalheight=\fh]{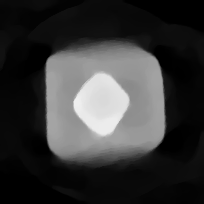} & \includegraphics[totalheight=\fh]{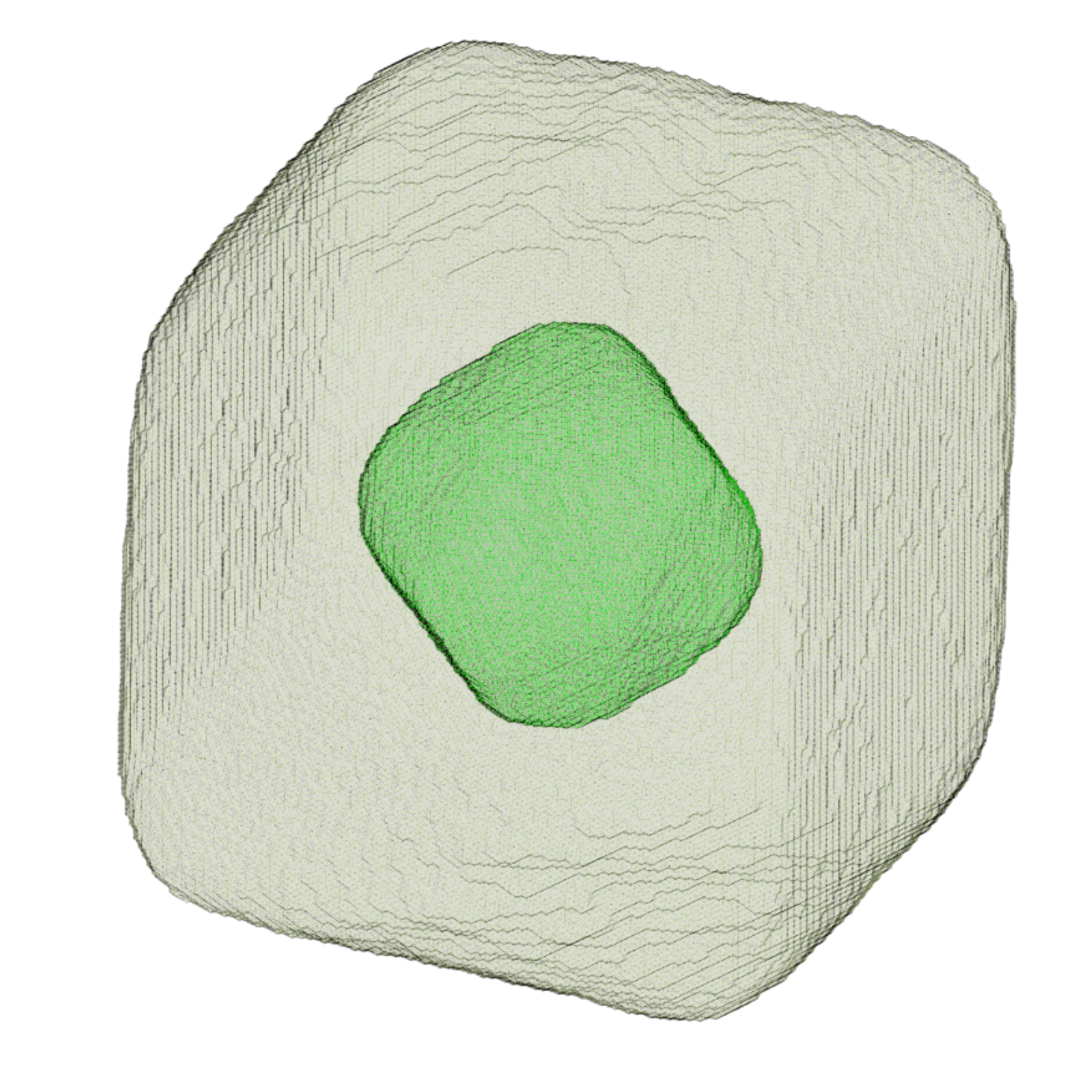} \\
\parbox[b][\fh][c]{\fw}{Proposed\\\small{(-$72^\circ\sim72^\circ$)}}  & \includegraphics[totalheight=\fh]{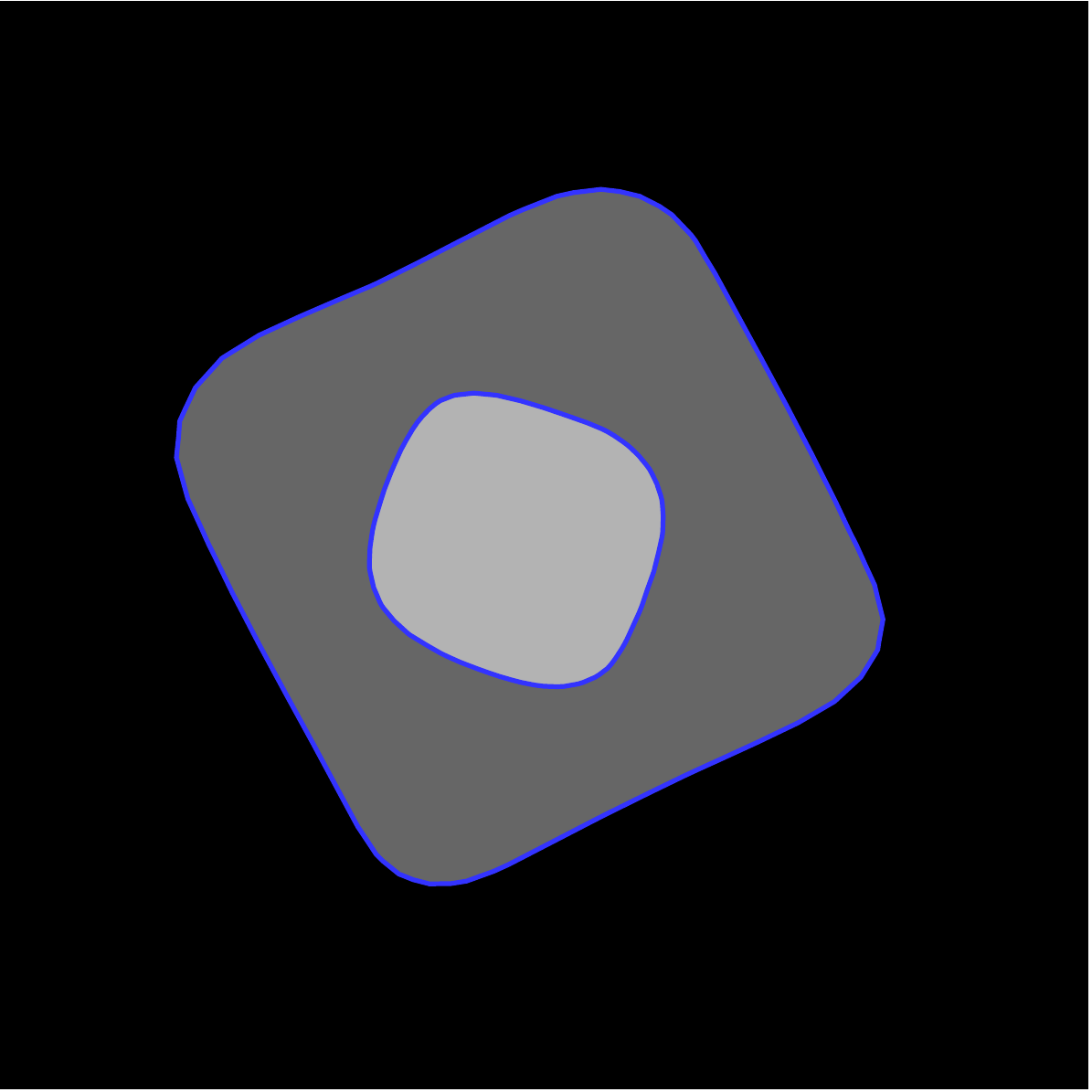} &\includegraphics[totalheight=\fh]{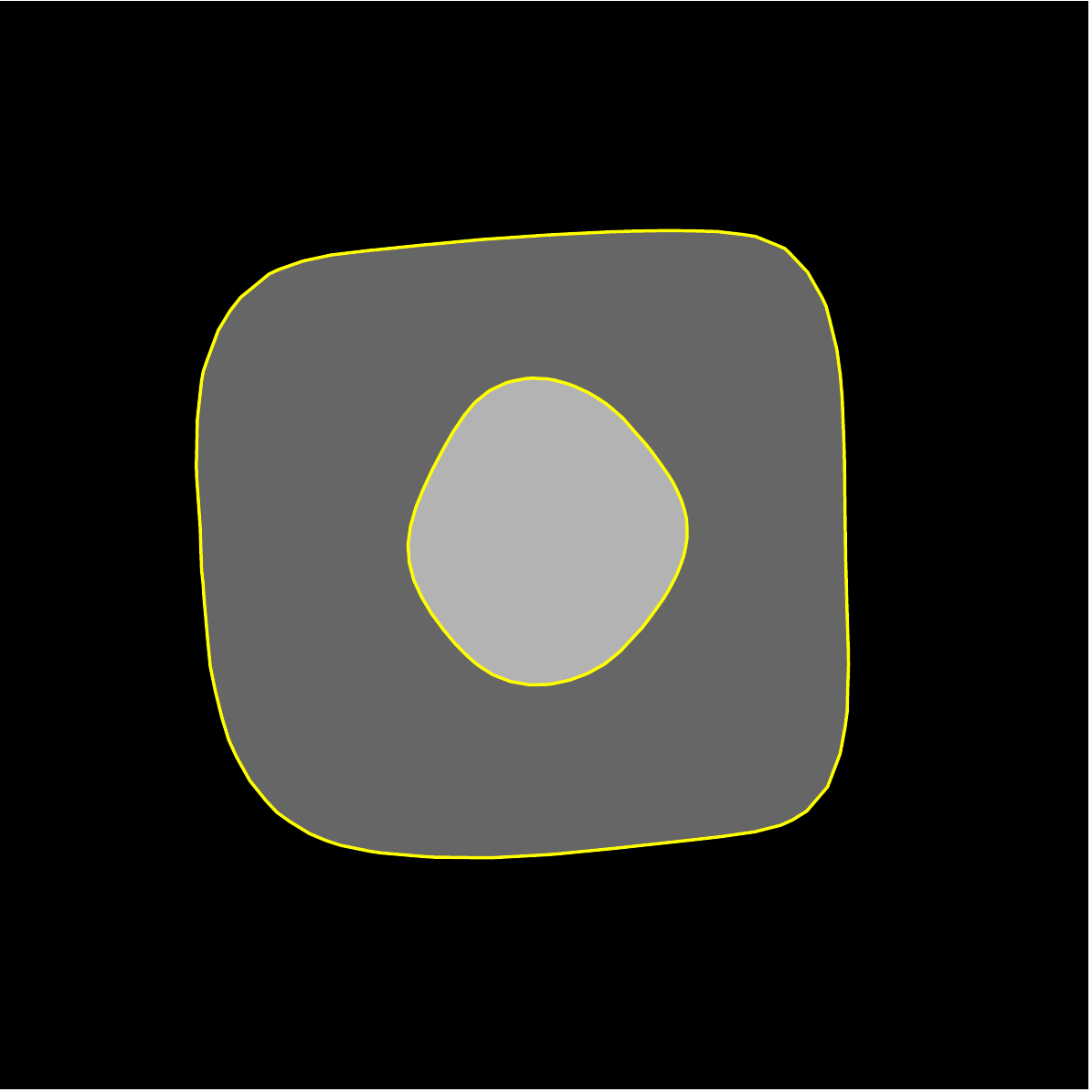} & \includegraphics[totalheight=\fh]{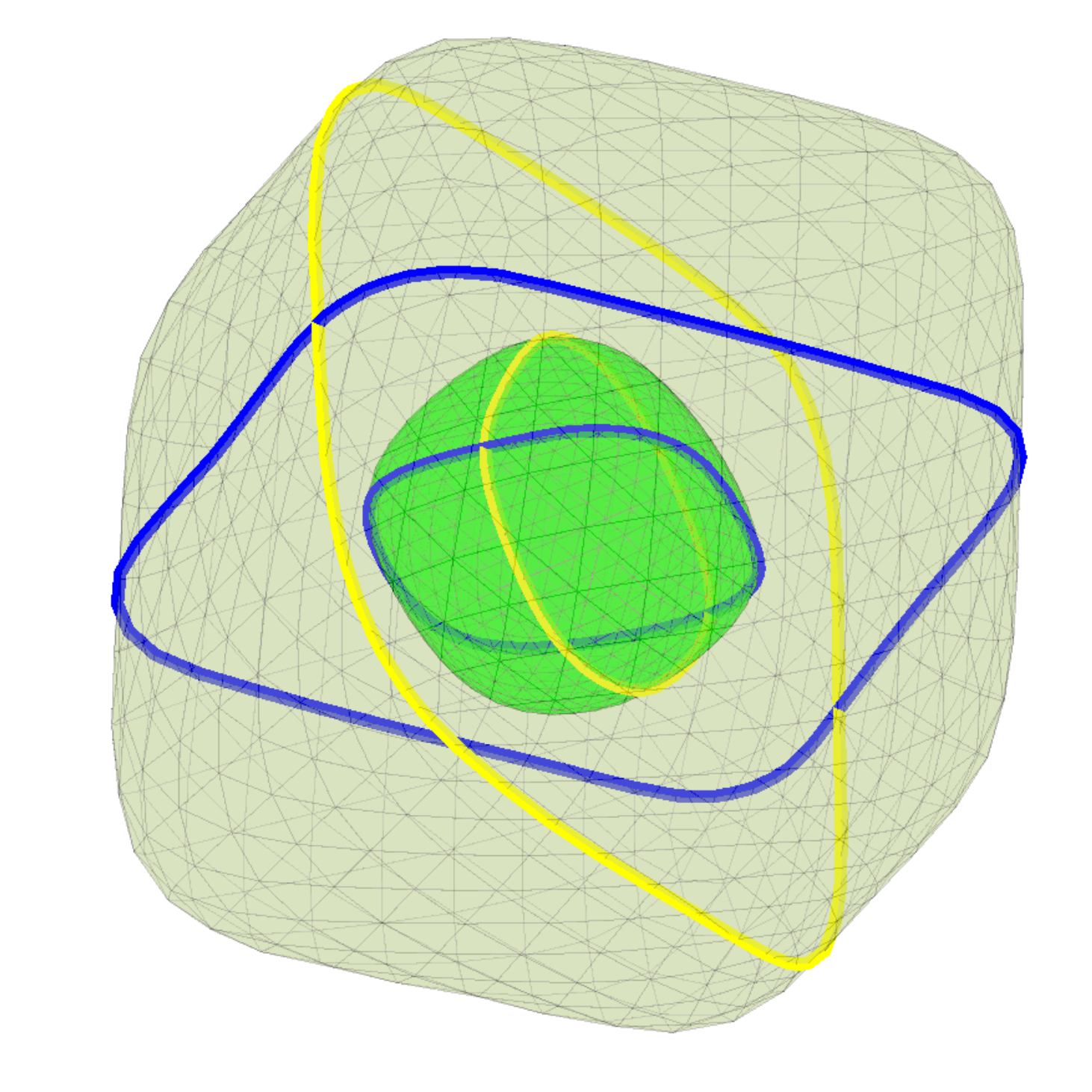} \\

\parbox[b][\fh][c]{\fw}{SIRT+Isosurface\\\small{(-$18^\circ\sim18^\circ$)}} & \includegraphics[totalheight=\fh]{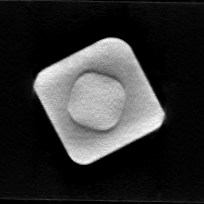} & \includegraphics[totalheight=\fh]{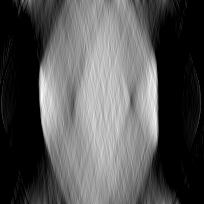} & \includegraphics[totalheight=\fh]{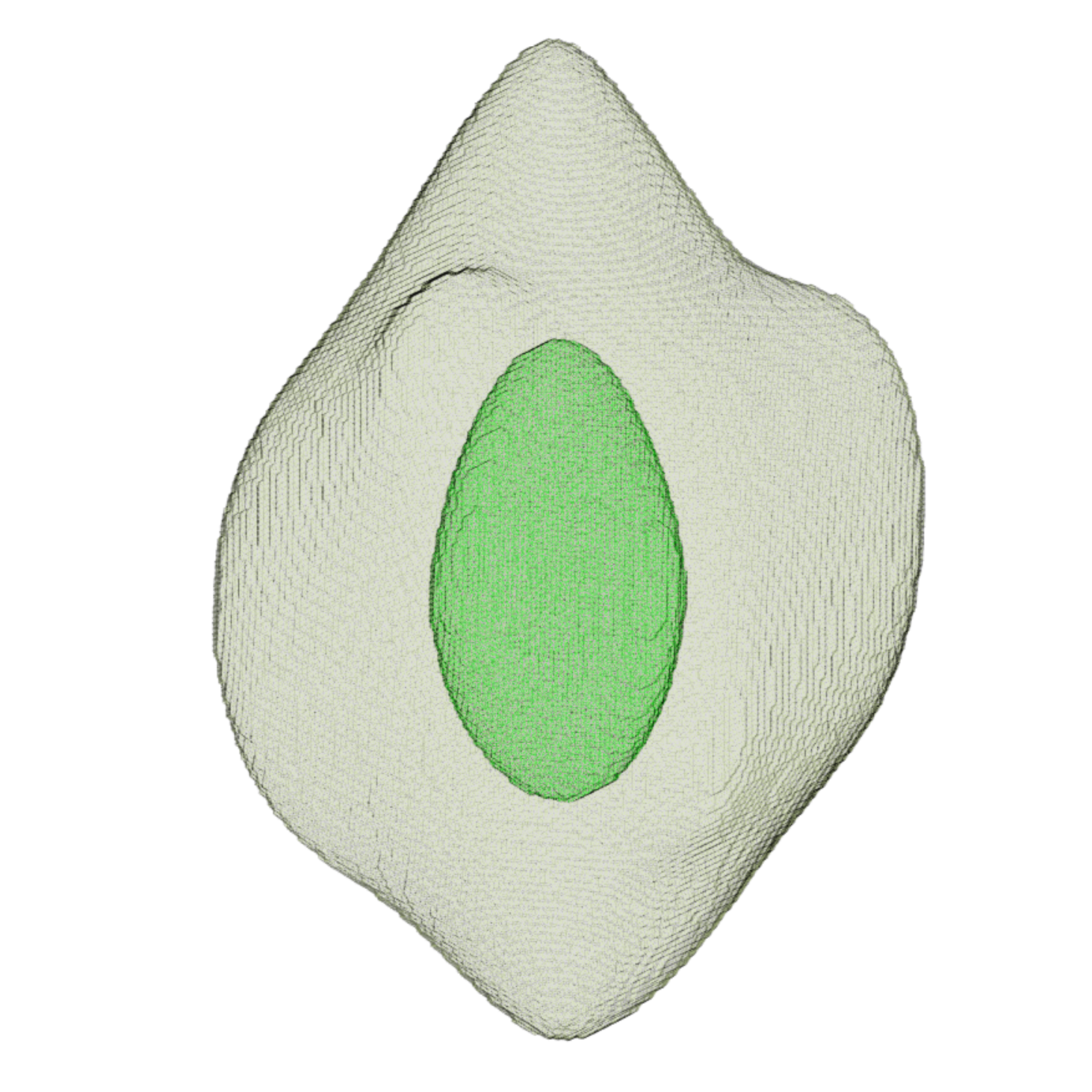}  \\
\parbox[b][\fh][c]{\fw}{TV+Isosurface\\\small{(-$18^\circ\sim18^\circ$)}} & \includegraphics[totalheight=\fh]{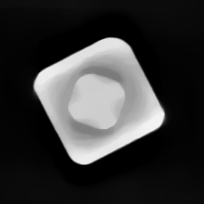} & \includegraphics[totalheight=\fh]{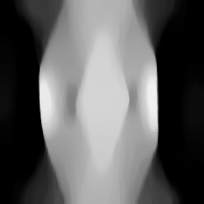} & \includegraphics[totalheight=\fh]{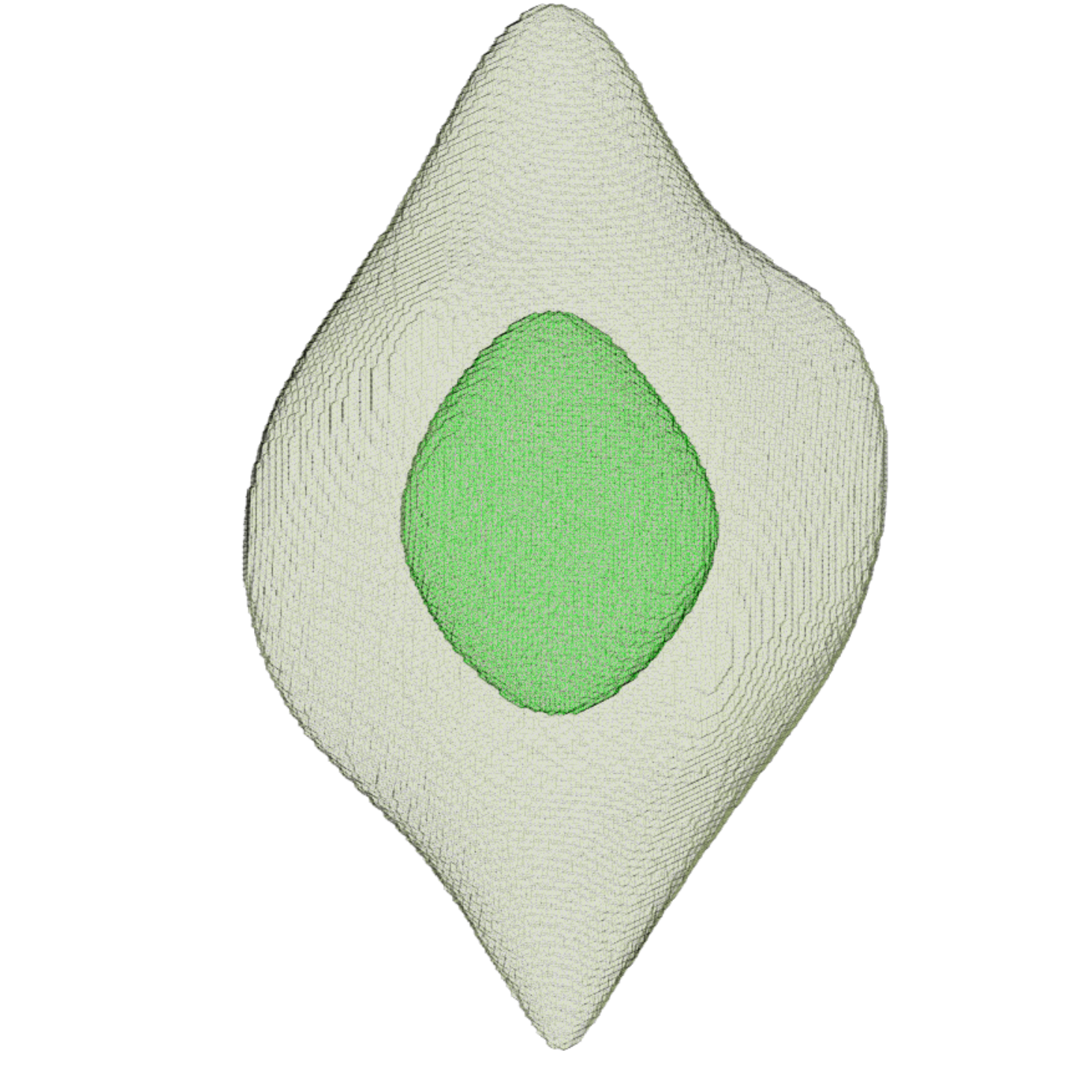}  \\
\parbox[b][\fh][c]{\fw}{Proposed\\\small{(-$18^\circ\sim18^\circ$)}}  & \includegraphics[totalheight=\fh]{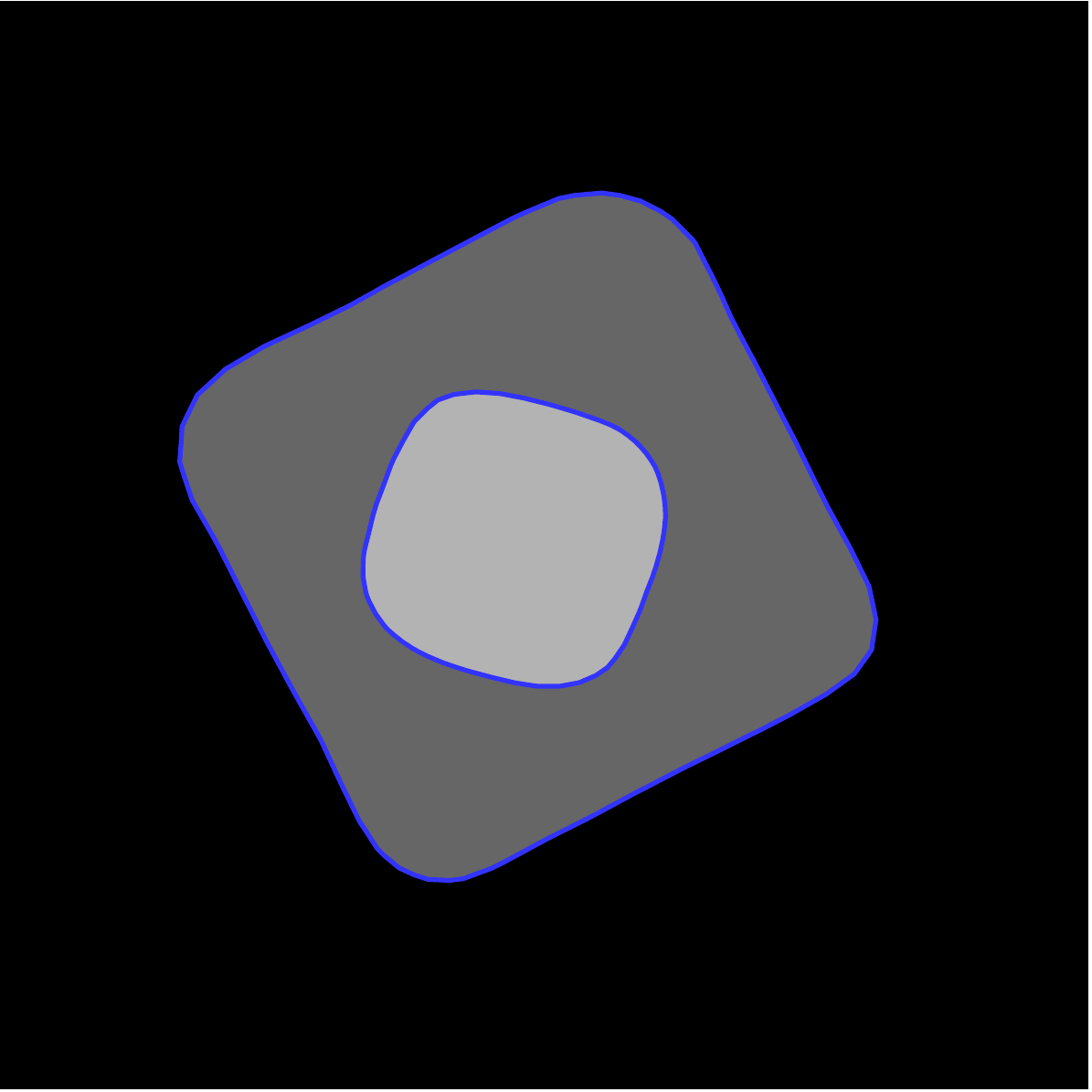} &\includegraphics[totalheight=\fh]{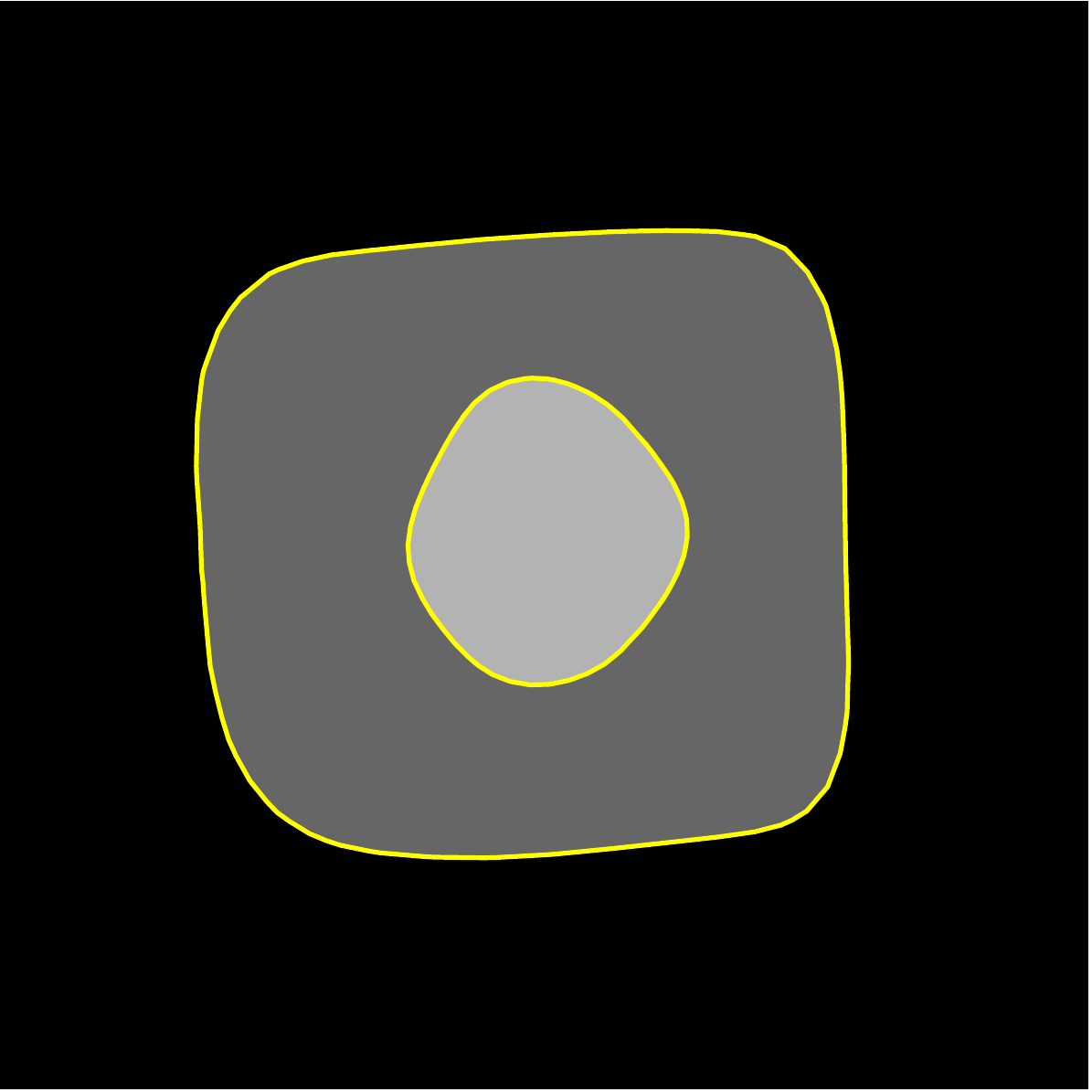} & \includegraphics[totalheight=\fh]{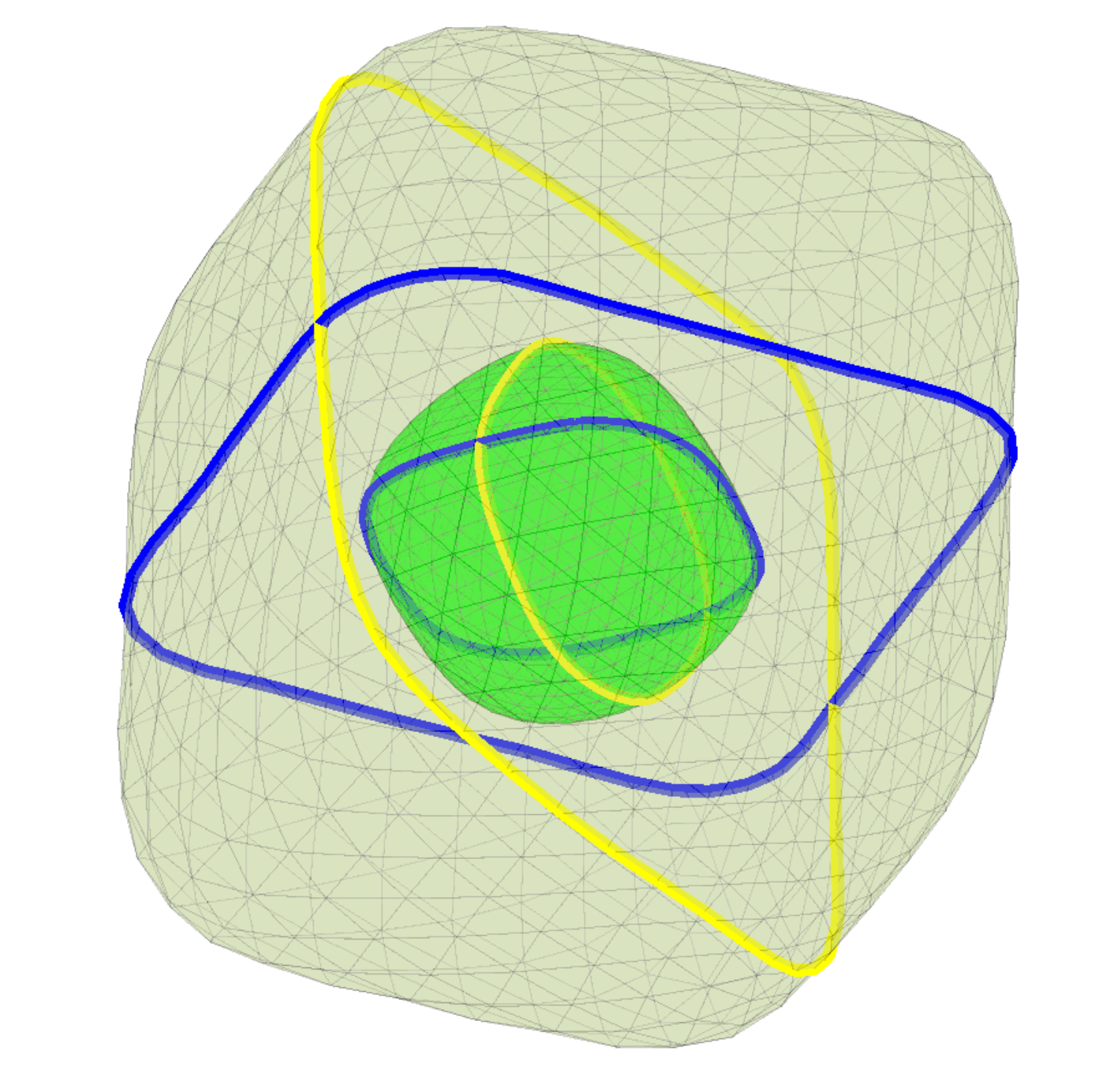}
\end{tabular}
\caption{Reconstruction results on nano cube data. The first and the 4th rows show the reconstruction results by SIRT and from projections with tilt angles from -72$^\circ$ to 72$^\circ$ and -18$^\circ$ to 18$^\circ$, respectively, where two central slices are visualized in the first and second column. The 2nd and the 5th rows show the reconstruction results by TV. The last column shows the extracted mesh from the reconstructed image by SIRT or TV with some post-processing procedures. The third and the last row show our direct shape estimation results, where two central slices are visualized in blue and yellow. The results on the second and the fourth row appear to be the same, but they are slightly different.}
\label{fig:nanocube}
\end{center}
\end{figure*}

\section{Conclusion and Discussion}

In this paper, we have developed a forward projector to map a triangular mesh onto the projections domain which is differentiable with respect to the mesh vertices and the attenuation values. Based on our differentiable forward model, we have suggested an optimization-based shape estimation method from projections. Our model should be chosen for reconstructing homogeneous objects with relatively simple geometry, but in situations where the data foundation is so noisy or limited that other methods will not allow a reconstruction. While a conventional approach most often consists of the two steps of image reconstruction followed by a segmentation to obtain a surface, the proposed method directly yields the surface, which among others allows imposing a shape prior directly on the reconstructed shapes. Our experiments on synthetic data show that the proposed method is robust to noise when reconstructing single objects. The experiments on the electron tomography data show how our method is robust even when the range of tilt angles is highly limited, compared with SIRT and TV. In this challenging case, the conventional image reconstructions by SIRT and TV are highly degraded and these degenerate results propagate errors to the surface estimation from the reconstructed images. Our method can, however, capture the overall shapes well with the regularization term which can impose a smoothness prior directly on the object shapes.

To discuss the limitation of the proposed method, we divide our contributions into two parts: the differentiable forward model and the shape estimation method. The proposed forward model estimates the displacement direction of each vertex for minimizng the objective function. These displacements will only allow change in shape and not collisions of splitting of the mesh that would require topological changes. Handling such topological changes and collisions is itself an active research topic in mesh deformation in computer graphics or computer vision. Investigating how to handle topology changes for a more general mesh adaption and deformation is for future research.

The current differentiable forward model has only been tested on parallel beam geometry where rays are perpendicular to the detector plane. This can be extended to a more general setting such as the cone beam geometry~\citep{buzug2008computed}, but we have not investigated the effect of that, as electron tomography data are based on parallel beam geometry. 

As a potential application, the proposed method can be useful for estimating the geometry precisely, e.g. for registering the reconstructed object to a 3D CAD model. Another future application can be dynamic tomography where we aim to analyze the objects which may change during scanning.

\section*{Acknowledgments}
This work is funded by EU Horizon 2020 MSCA Innovative Training Network MUMMERING Grant Number 765604.

\bibliographystyle{model2-names}
{\small
\bibliography{ctdr}

\begin{thebibliography}{33}
\expandafter\ifx\csname natexlab\endcsname\relax\def\natexlab#1{#1}\fi
\providecommand{\url}[1]{\texttt{#1}}
\providecommand{\href}[2]{#2}
\providecommand{\path}[1]{#1}
\providecommand{\DOIprefix}{doi:}
\providecommand{\ArXivprefix}{arXiv:}
\providecommand{\URLprefix}{URL: }
\providecommand{\Pubmedprefix}{pmid:}
\providecommand{\doi}[1]{\href{http://dx.doi.org/#1}{\path{#1}}}
\providecommand{\Pubmed}[1]{\href{pmid:#1}{\path{#1}}}
\providecommand{\bibinfo}[2]{#2}
\ifx\xfnm\undefined \def\xfnm[#1]{\unskip,\space#1}\fi
\bibitem[{Aghasi et~al.(2011)Aghasi, Kilmer and Miller}]{aghasi2011parametric}
\bibinfo{author}{Aghasi\xfnm[ A.]}, \bibinfo{author}{Kilmer\xfnm[ M.]},
  \bibinfo{author}{Miller\xfnm[ E.L.]}.
\newblock \bibinfo{title}{Parametric {{Level Set Methods}} for {{Inverse
  Problems}}}.
\newblock \bibinfo{journal}{SIAM J Imaging Sci}
  \bibinfo{year}{2011};\bibinfo{volume}{4}(\bibinfo{number}{2}).
\bibitem[{Alvino and Yezzi(2004)}]{alvino2004tomographic}
\bibinfo{author}{Alvino\xfnm[ C.V.]}, \bibinfo{author}{Yezzi\xfnm[ A.J.]}.
\newblock \bibinfo{title}{Tomographic reconstruction of piecewise smooth
  images}.
\newblock In: \bibinfo{booktitle}{{{IEEE Conference on Computer Vision and
  Pattern Recognition}}}. \bibinfo{year}{2004}. .
\bibitem[{Andersen and Kak(1984)}]{andersen1984simultaneous}
\bibinfo{author}{Andersen\xfnm[ A.H.]}, \bibinfo{author}{Kak\xfnm[ A.C.]}.
\newblock \bibinfo{title}{Simultaneous {{Algebraic Reconstruction Technique}}
  ({{SART}}): {{A}} superior implementation of the {{ART}} algorithm}.
\newblock \bibinfo{journal}{Ultrason Imaging}
  \bibinfo{year}{1984};\bibinfo{volume}{6}(\bibinfo{number}{1}).
\bibitem[{Attene(2010)}]{attene2010lightweight}
\bibinfo{author}{Attene\xfnm[ M.]}.
\newblock \bibinfo{title}{A lightweight approach to repairing digitized polygon
  meshes}.
\newblock \bibinfo{journal}{Vis Comput}
  \bibinfo{year}{2010};\bibinfo{volume}{26}(\bibinfo{number}{11}).
\bibitem[{Bonin et~al.(2002)Bonin, Chalmond and
  Lavayssi{\`e}re}]{bonin2002montecarlo}
\bibinfo{author}{Bonin\xfnm[ A.]}, \bibinfo{author}{Chalmond\xfnm[ B.]},
  \bibinfo{author}{Lavayssi{\`e}re\xfnm[ B.]}.
\newblock \bibinfo{title}{Monte-{{Carlo}} simulation of industrial radiography
  images and experimental designs}.
\newblock \bibinfo{journal}{NDT \& E International}
  \bibinfo{year}{2002};\bibinfo{volume}{35}(\bibinfo{number}{8}).
\bibitem[{Buzug(2008)}]{buzug2008computed}
\bibinfo{author}{Buzug\xfnm[ T.M.]}.
\newblock \bibinfo{title}{Computed Tomography from Photon Statistics to Modern
  Cone Beam {{CT}}}.
\newblock \bibinfo{publisher}{{Springer}}, \bibinfo{year}{2008}.
\bibitem[{Chambolle and Pock(2011)}]{chambolle2011firstorder}
\bibinfo{author}{Chambolle\xfnm[ A.]}, \bibinfo{author}{Pock\xfnm[ T.]}.
\newblock \bibinfo{title}{A {{First}}-{{Order Primal}}-{{Dual Algorithm}} for
  {{Convex Problems}} with {{Applications}} to {{Imaging}}}.
\newblock \bibinfo{journal}{J Math Imaging Vis}
  \bibinfo{year}{2011};\bibinfo{volume}{40}(\bibinfo{number}{1}).
\bibitem[{Chen et~al.(2019)Chen, Gao, Ling, Smith, Lehtinen, Jacobson and
  Fidler}]{chen2019learning}
\bibinfo{author}{Chen\xfnm[ W.]}, \bibinfo{author}{Gao\xfnm[ J.]},
  \bibinfo{author}{Ling\xfnm[ H.]}, \bibinfo{author}{Smith\xfnm[ E.J.]},
  \bibinfo{author}{Lehtinen\xfnm[ J.]}, \bibinfo{author}{Jacobson\xfnm[ A.]},
  \bibinfo{author}{Fidler\xfnm[ S.]}.
\newblock \bibinfo{title}{Learning to {{Predict 3D Objects}} with an
  {{Interpolation}}-based {{Differentiable Renderer}}}.
\newblock In: \bibinfo{booktitle}{{{Neural Information Processing Systems}}}.
  \bibinfo{year}{2019}. .
\bibitem[{Dahl et~al.(2018)Dahl, Dahl and Hansen}]{dahl2018computing}
\bibinfo{author}{Dahl\xfnm[ V.A.]}, \bibinfo{author}{Dahl\xfnm[ A.B.]},
  \bibinfo{author}{Hansen\xfnm[ P.C.]}.
\newblock \bibinfo{title}{Computing segmentations directly from x-ray
  projection data via parametric deformable curves}.
\newblock \bibinfo{journal}{Meas Sci Technol}
  \bibinfo{year}{2018};\bibinfo{volume}{29}(\bibinfo{number}{1}).
\bibitem[{Elangovan and Whitaker(2001)}]{elangovan2001sinograms}
\bibinfo{author}{Elangovan\xfnm[ V.]}, \bibinfo{author}{Whitaker\xfnm[ R.T.]}.
\newblock \bibinfo{title}{From {{Sinograms}} to {{Surfaces}}: {{A Direct
  Approach}} to the {{Segmentation}} of {{Tomographic Data}}}.
\newblock In: \bibinfo{booktitle}{Medical {{Image Computing}} and
  {{Computer}}-{{Assisted Intervention}} ({{MICCAI}})}. \bibinfo{year}{2001}. .
\bibitem[{Eliasof et~al.(2020)Eliasof, Sharf and
  Treister}]{eliasof2020multimodal}
\bibinfo{author}{Eliasof\xfnm[ M.]}, \bibinfo{author}{Sharf\xfnm[ A.]},
  \bibinfo{author}{Treister\xfnm[ E.]}.
\newblock \bibinfo{title}{Multi-modal {{3D}} shape reconstruction under
  calibration uncertainty using parametric level set methods}.
\newblock \bibinfo{journal}{SIAM J Imaging Sci}
  \bibinfo{year}{2020};\bibinfo{volume}{13}(\bibinfo{number}{1}).
\bibitem[{Freud et~al.(2006)Freud, Duvauchelle, L{\'e}tang and
  Babot}]{freud2006fast}
\bibinfo{author}{Freud\xfnm[ N.]}, \bibinfo{author}{Duvauchelle\xfnm[ P.]},
  \bibinfo{author}{L{\'e}tang\xfnm[ J.M.]}, \bibinfo{author}{Babot\xfnm[ D.]}.
\newblock \bibinfo{title}{Fast and robust ray casting algorithms for virtual
  {{X}}-ray imaging}.
\newblock \bibinfo{journal}{Nuclear Instruments and Methods in Physics Research
  Section B: Beam Interactions with Materials and Atoms}
  \bibinfo{year}{2006};\bibinfo{volume}{248}(\bibinfo{number}{1}).
\bibitem[{Gadelha et~al.(2019)Gadelha, Wang and Maji}]{gadelha2019shape}
\bibinfo{author}{Gadelha\xfnm[ M.]}, \bibinfo{author}{Wang\xfnm[ R.]},
  \bibinfo{author}{Maji\xfnm[ S.]}.
\newblock \bibinfo{title}{Shape {{Reconstruction Using Differentiable
  Projections}} and {{Deep Priors}}}.
\newblock In: \bibinfo{booktitle}{{{International Conference on Computer
  Vision}}}. \bibinfo{year}{2019}. .
\bibitem[{Huang et~al.(2018)Huang, Su and Guibas}]{huang2018robust}
\bibinfo{author}{Huang\xfnm[ J.]}, \bibinfo{author}{Su\xfnm[ H.]},
  \bibinfo{author}{Guibas\xfnm[ L.]}.
\newblock \bibinfo{title}{Robust {{Watertight Manifold Surface Generation
  Method}} for {{ShapeNet Models}}}.
\newblock \bibinfo{journal}{ArXiv180201698 Cs} \bibinfo{year}{2018};.
\bibitem[{Kadu et~al.(2018)Kadu, {van Leeuwen} and
  Batenburg}]{kadu2018parametric}
\bibinfo{author}{Kadu\xfnm[ A.]}, \bibinfo{author}{{van Leeuwen}\xfnm[ T.]},
  \bibinfo{author}{Batenburg\xfnm[ K.J.]}.
\newblock \bibinfo{title}{A parametric level-set method for partially discrete
  tomography}.
\newblock In: \bibinfo{booktitle}{International {{Conference}} on {{Discrete
  Geometry}} for {{Computer Imagery}}}. \bibinfo{year}{2018}. .
\bibitem[{Kass et~al.(1988)Kass, Witkin and Terzopoulos}]{kass1988snakes}
\bibinfo{author}{Kass\xfnm[ M.]}, \bibinfo{author}{Witkin\xfnm[ A.]},
  \bibinfo{author}{Terzopoulos\xfnm[ D.]}.
\newblock \bibinfo{title}{Snakes: {{Active}} contour models}.
\newblock \bibinfo{journal}{Int J Comput Vis}
  \bibinfo{year}{1988};\bibinfo{volume}{1}(\bibinfo{number}{4}).
\bibitem[{Kato et~al.(2018)Kato, Ushiku and Harada}]{kato2018neural}
\bibinfo{author}{Kato\xfnm[ H.]}, \bibinfo{author}{Ushiku\xfnm[ Y.]},
  \bibinfo{author}{Harada\xfnm[ T.]}.
\newblock \bibinfo{title}{Neural {{3D Mesh Renderer}}}.
\newblock In: \bibinfo{booktitle}{{{IEEE Conference on Computer Vision and
  Pattern Recognition}}}. \bibinfo{year}{2018}. .
\bibitem[{Kingma and Ba(2015)}]{kingma2015adam}
\bibinfo{author}{Kingma\xfnm[ D.]}, \bibinfo{author}{Ba\xfnm[ J.]}.
\newblock \bibinfo{title}{Adam: {{A}} method for stochastic optimization}.
\newblock In: \bibinfo{booktitle}{International {{Conference}} on {{Learning
  Representations}}}. \bibinfo{year}{2015}. .
\bibitem[{Lauterbach et~al.(2009)Lauterbach, Garland, Sengupta, Luebke and
  Manocha}]{lauterbach2009fast}
\bibinfo{author}{Lauterbach\xfnm[ C.]}, \bibinfo{author}{Garland\xfnm[ M.]},
  \bibinfo{author}{Sengupta\xfnm[ S.]}, \bibinfo{author}{Luebke\xfnm[ D.P.]},
  \bibinfo{author}{Manocha\xfnm[ D.]}.
\newblock \bibinfo{title}{Fast {{BVH Construction}} on {{GPUs}}}.
\newblock \bibinfo{journal}{Comput Graph Forum}
  \bibinfo{year}{2009};\bibinfo{volume}{28}(\bibinfo{number}{2}).
\bibitem[{Liu et~al.(2019)Liu, Chen, Li and Li}]{liu2019soft}
\bibinfo{author}{Liu\xfnm[ S.]}, \bibinfo{author}{Chen\xfnm[ W.]},
  \bibinfo{author}{Li\xfnm[ T.]}, \bibinfo{author}{Li\xfnm[ H.]}.
\newblock \bibinfo{title}{Soft {{Rasterizer}}: {{Differentiable Rendering}} for
  {{Unsupervised Single}}-{{View Mesh Reconstruction}}}.
\newblock In: \bibinfo{booktitle}{International Conference on Computer Vision}.
  \bibinfo{year}{2019}. .
\bibitem[{Loper and Black(2014)}]{loper2014opendr}
\bibinfo{author}{Loper\xfnm[ M.M.]}, \bibinfo{author}{Black\xfnm[ M.J.]}.
\newblock \bibinfo{title}{{{OpenDR}}: {{An Approximate Differentiable
  Renderer}}}.
\newblock In: \bibinfo{booktitle}{European {{Conference}} on {{Computer
  Vision}}}. \bibinfo{year}{2014}. .
\bibitem[{Marinovszki et~al.(2018)Marinovszki, Beenhouwer and
  Sijbers}]{marinovszki2018efficient}
\bibinfo{author}{Marinovszki\xfnm[ {\'A}.]}, \bibinfo{author}{Beenhouwer\xfnm[
  J.D.]}, \bibinfo{author}{Sijbers\xfnm[ J.]}.
\newblock \bibinfo{title}{An efficient {{CAD}} projector for {{X}}-ray
  projection based {{3D}} inspection with the {{ASTRA Toolbox}}}.
\newblock In: \bibinfo{booktitle}{Conference on {{Industrial Computed
  Tomography}}}. \bibinfo{year}{2018}. .
\bibitem[{Midgley and Weyland(2003)}]{midgley20033d}
\bibinfo{author}{Midgley\xfnm[ P.]}, \bibinfo{author}{Weyland\xfnm[ M.]}.
\newblock \bibinfo{title}{3d electron microscopy in the physical sciences: the
  development of z-contrast and eftem tomography}.
\newblock \bibinfo{journal}{Ultramicroscopy}
  \bibinfo{year}{2003};\bibinfo{volume}{96}(\bibinfo{number}{3}):\bibinfo{pages}{413--431}.
\bibitem[{Mueller and Siltanen(2012)}]{mueller2012linear}
\bibinfo{author}{Mueller\xfnm[ J.L.]}, \bibinfo{author}{Siltanen\xfnm[ S.]}.
\newblock \bibinfo{title}{Linear and Nonlinear Inverse Problems with Practical
  Applications}.
\newblock volume~\bibinfo{volume}{10}.
\newblock \bibinfo{publisher}{{Society for Industrial and Applied
  Mathematics}}, \bibinfo{year}{2012}.
\bibitem[{Mumford and Shah(1989)}]{mumford1989optimal}
\bibinfo{author}{Mumford\xfnm[ D.]}, \bibinfo{author}{Shah\xfnm[ J.]}.
\newblock \bibinfo{title}{Optimal approximations by piecewise smooth functions
  and associated variational problems}.
\newblock \bibinfo{journal}{Commun Pure Appl Math}
  \bibinfo{year}{1989};\bibinfo{volume}{42}(\bibinfo{number}{5}).
\bibitem[{Osher and Fedkiw(2004)}]{osher2004level}
\bibinfo{author}{Osher\xfnm[ S.]}, \bibinfo{author}{Fedkiw\xfnm[ R.]}.
\newblock \bibinfo{title}{Level set methods and dynamic implicit surfaces}.
\newblock \bibinfo{journal}{Appl Mech Rev}
  \bibinfo{year}{2004};\bibinfo{volume}{57}(\bibinfo{number}{3}).
\bibitem[{Paszke et~al.(2019)Paszke, Gross, Massa, Lerer, Bradbury, Chanan,
  Killeen, Lin, Gimelshein, Antiga, Desmaison, K{\"o}pf, Yang, DeVito, Raison,
  Tejani, Chilamkurthy, Steiner, Fang, Bai and Chintala}]{paszke2019pytorch}
\bibinfo{author}{Paszke\xfnm[ A.]}, \bibinfo{author}{Gross\xfnm[ S.]},
  \bibinfo{author}{Massa\xfnm[ F.]}, \bibinfo{author}{Lerer\xfnm[ A.]},
  \bibinfo{author}{Bradbury\xfnm[ J.]}, \bibinfo{author}{Chanan\xfnm[ G.]},
  \bibinfo{author}{Killeen\xfnm[ T.]}, \bibinfo{author}{Lin\xfnm[ Z.]},
  \bibinfo{author}{Gimelshein\xfnm[ N.]}, \bibinfo{author}{Antiga\xfnm[ L.]},
  \bibinfo{author}{Desmaison\xfnm[ A.]}, \bibinfo{author}{K{\"o}pf\xfnm[ A.]},
  \bibinfo{author}{Yang\xfnm[ E.]}, \bibinfo{author}{DeVito\xfnm[ Z.]},
  \bibinfo{author}{Raison\xfnm[ M.]}, \bibinfo{author}{Tejani\xfnm[ A.]},
  \bibinfo{author}{Chilamkurthy\xfnm[ S.]}, \bibinfo{author}{Steiner\xfnm[
  B.]}, \bibinfo{author}{Fang\xfnm[ L.]}, \bibinfo{author}{Bai\xfnm[ J.]},
  \bibinfo{author}{Chintala\xfnm[ S.]}.
\newblock \bibinfo{title}{{{PyTorch}}: {{An}} imperative style,
  high-performance deep learning library}.
\newblock In: \bibinfo{booktitle}{{{Neural Information Processing Systems}}}.
  \bibinfo{year}{2019}. .
\bibitem[{Roelandts et~al.(2014)Roelandts, Batenburg, {den Dekker} and
  Sijbers}]{roelandts2014reconstructed}
\bibinfo{author}{Roelandts\xfnm[ T.]}, \bibinfo{author}{Batenburg\xfnm[ K.J.]},
  \bibinfo{author}{{den Dekker}\xfnm[ A.J.]}, \bibinfo{author}{Sijbers\xfnm[
  J.]}.
\newblock \bibinfo{title}{The reconstructed residual error: {{A}} novel
  segmentation evaluation measure for reconstructed images in tomography}.
\newblock \bibinfo{journal}{Comput Vis Image Underst}
  \bibinfo{year}{2014};\bibinfo{volume}{126}.
\bibitem[{Sidky et~al.(2012)Sidky, J{\o}rgensen and Pan}]{sidky2012convex}
\bibinfo{author}{Sidky\xfnm[ E.Y.]}, \bibinfo{author}{J{\o}rgensen\xfnm[
  J.H.]}, \bibinfo{author}{Pan\xfnm[ X.]}.
\newblock \bibinfo{title}{Convex optimization problem prototyping for image
  reconstruction in computed tomography with the
  {{Chambolle}}\textendash{{Pock}} algorithm}.
\newblock \bibinfo{journal}{Phys Med Biol}
  \bibinfo{year}{2012};\bibinfo{volume}{57}(\bibinfo{number}{10}).
\bibitem[{Sujar et~al.(2017)Sujar, Meuleman, Villard, Garcia and
  Vidal}]{sujar2017gvirtualxray}
\bibinfo{author}{Sujar\xfnm[ A.]}, \bibinfo{author}{Meuleman\xfnm[ A.]},
  \bibinfo{author}{Villard\xfnm[ P.F.]}, \bibinfo{author}{Garcia\xfnm[ M.]},
  \bibinfo{author}{Vidal\xfnm[ F.]}.
\newblock \bibinfo{title}{{{gVirtualXRay}}: Virtual x-ray imaging library on
  {{GPU}}}.
\newblock In: \bibinfo{booktitle}{Computer {{Graphics}} and {{Visual
  Computing}} ({{CGVC}})}. \bibinfo{year}{2017}. .
\bibitem[{Vidal et~al.(2009)Vidal, Garnier, Freud, L{\'e}tang and
  John}]{vidal2009simulation}
\bibinfo{author}{Vidal\xfnm[ F.P.]}, \bibinfo{author}{Garnier\xfnm[ M.]},
  \bibinfo{author}{Freud\xfnm[ N.]}, \bibinfo{author}{L{\'e}tang\xfnm[ J.M.]},
  \bibinfo{author}{John\xfnm[ N.W.]}.
\newblock \bibinfo{title}{Simulation of {{X}}-ray attenuation on the {{GPU}}}.
\newblock In: \bibinfo{booktitle}{Proceedings of Theory and Practice of
  Computer Graphics}. \bibinfo{address}{{Cardiff, UK}}:
  \bibinfo{publisher}{{Eurographics Association}}; \bibinfo{year}{2009}. .
\bibitem[{Wang et~al.(2018)Wang, Zhang, Li, Fu, Liu and
  Jiang}]{wang2018pixel2mesh}
\bibinfo{author}{Wang\xfnm[ N.]}, \bibinfo{author}{Zhang\xfnm[ Y.]},
  \bibinfo{author}{Li\xfnm[ Z.]}, \bibinfo{author}{Fu\xfnm[ Y.]},
  \bibinfo{author}{Liu\xfnm[ W.]}, \bibinfo{author}{Jiang\xfnm[ Y.G.]}.
\newblock \bibinfo{title}{{{Pixel2Mesh}}: {{Generating 3D Mesh Models}} from
  {{Single RGB Images}}}.
\newblock In: \bibinfo{booktitle}{{{European Conference on Computer Vision}}}.
  \bibinfo{year}{2018}. .
\bibitem[{Whitaker and Elangovan(2002)}]{whitaker2002direct}
\bibinfo{author}{Whitaker\xfnm[ R.T.]}, \bibinfo{author}{Elangovan\xfnm[ V.]}.
\newblock \bibinfo{title}{A direct approach to estimating surfaces in
  tomographic data}.
\newblock \bibinfo{journal}{Med Image Anal}
  \bibinfo{year}{2002};\bibinfo{volume}{6}(\bibinfo{number}{3}).

\end{thebibliography}
}
\end{document}